%% file: main.tex
\documentclass[sigconf,nonacm]{acmart}
\AtBeginDocument{%
  }

\setcopyright{cc}
\setcctype{by}
\usepackage{xspace}
\usepackage[utf8]{inputenc}
\usepackage[dvipsnames]{xcolor}
\usepackage{array}
\usepackage{booktabs}
\usepackage{makecell}
\usepackage{listings}
\usepackage{float}
\newcommand{\system}{\textsc{Omakase}\xspace}

\begin{document}

% \title{Omakase: project context-aware agent}
% \title{Omakase tailored research literature assistance}
\title{\system{}: proactive assistance with actionable suggestions for evolving scientific research projects}

\author{Pao Siangliulue}
\email{paos@allenai.org}
\affiliation{
    \institution{Allen Institute for AI}
    \country{USA}
}

\author{Jonathan Bragg}
\email{jbragg@allenai.org}
\affiliation{
    \institution{Allen Institute for AI}
    \country{USA}
}

\author{Doug Downey}
\email{dougd@allenai.org}
\affiliation{
    \institution{Allen Institute for AI}
    \country{USA}
}

\author{Joseph Chee Chang}
\email{josephc@allenai.org}
\affiliation{
    \institution{Allen Institute for AI}
    \country{USA}
}

\author{Daniel S. Weld}
\email{danw@allenai.org}
\affiliation{
  \institution{Allen Institute for AI}
  \country{USA}
}

\renewcommand{\shortauthors}{Siangliulue et al.}

\input{sections/00_abstract}

%% [Commented out for arXiv preprint]
%\begin{CCSXML}
%<ccs2012>
%</ccs2012>
%\end{CCSXML}

%% \ccsdesc[500]{Do Not Use This Code~Generate the Correct Terms for Your Paper}
%% \ccsdesc[300]{Do Not Use This Code~Generate the Correct Terms for Your Paper}
%% \ccsdesc{Do Not Use This Code~Generate the Correct Terms for Your Paper}
%% \ccsdesc[100]{Do Not Use This Code~Generate the Correct Terms for Your Paper}
\begin{teaserfigure}
  \includegraphics[width=\textwidth]{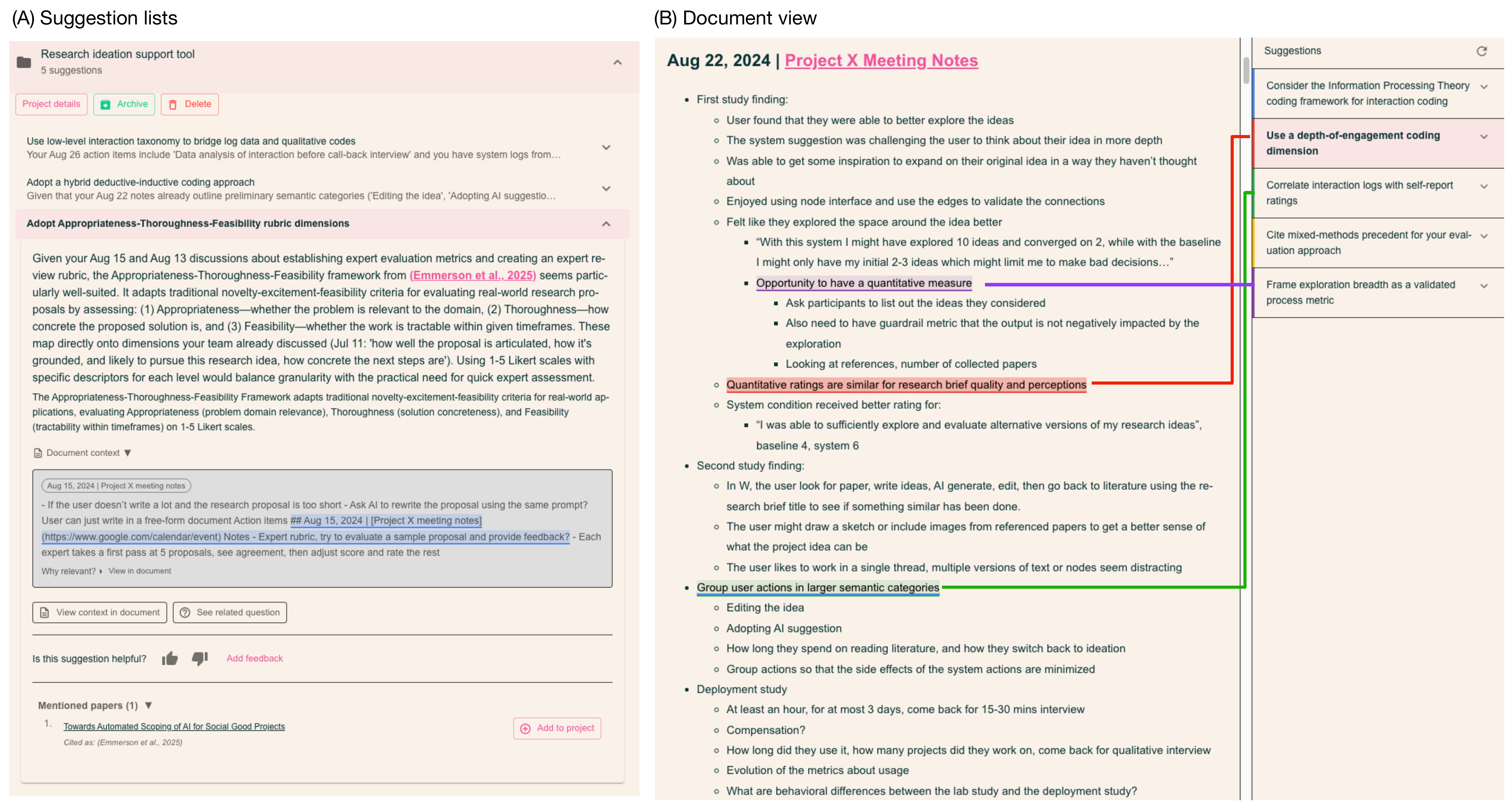}
  \caption{\system{} regularly sends literature-based actionable suggestions to help with a research project based on user's research project document. Once new suggestions are available, the user received a notification email which links to \system{} Dashboard with a list of actionable suggestions (A). The user can view suggestions, the relevant context from their document and the question answer from a deep research system that the suggestion was derived from. The document view (B) allows the user to explore the relevant parts of the document related to specific suggestions.}
  % \caption{After the user submit their research project document, \system{} parses the document details and presents inferred information about the project on the Project details page (A1). The user can edit the inferred project state (A3) if \system{} misinterpret. \system{} shows a list of questions for a literature review deep research system that would be helpful for the user(A4). The user can add and track questions. There is also a list of papers initially extracted from the document (A4); each paper comes with editable explanation of how the paper is relevant to the project. \system{} regularly send user actionable suggestions based on the information in the document. The user can set the frequency of these suggestions or get new suggestions manually (A2). Once new suggestions are available, the user received a notification email which links to \system{} Dashboard with a list of actionable suggestions (B). The user can view suggestions, the relevant context from their document (B6) and the question answer from a deep research system that the suggestion was derived from (B7). They can  provide feedback using the thumbs-up and down mechanism and free-form text (B8). Finally, the document view (C) allows the user to explore the relevant parts of the document related to specific suggestions.}
  \Description{An overview of the Omakase system showing how it monitors project documents and proactively delivers literature-grounded, actionable suggestions.}
  \label{fig:omakase-system}
\end{teaserfigure}

%% [Commented out for arXiv preprint]
%% \received{20 February 2007}
%% \received[revised]{12 March 2009}
%% \received[accepted]{5 June 2009}

\maketitle

\input{sections/01_introduction}
\input{sections/02_related_work}
\input{sections/03_deployment_study}
\input{sections/05_system}
\input{sections/06_suggestions_evaluation}

\input{sections/07_discussion}
\input{sections/08_conclusion}
% \input{sections/08_acknowledgements}

% \begin{acks}
% \end{acks}

\bibliographystyle{ACM-Reference-Format}
\bibliography{omakase}

\appendix
\input{sections/10_appendix}
% \appendix

\end{document}

%% file: sections/00_abstract.tex
\begin{abstract}
As AI agents become increasingly capable of complex knowledge tasks, the lack of context limits their capability to proactively reason about a user’s latent needs throughout a long evolving project. In scientific research, many researchers still manually query a deep research system and compress their rich project contexts into short, targeted queries. Further, a deep research system produces exhaustive reports, making it difficult to identify concrete actions. To explore the opportunities of research assistants that are proactive throughout a research project, we conducted several studies ($N=42$) with a technology probe and an iterative prototype. The latest iteration of our system, \system{}, is a research assistant that monitors a user’s project documents to infer timely queries to a deep research system. \system{} then distills long reports into suggestions contextualized to their evolving projects. Our evaluations showed that participants found the generated queries to be useful and timely, and rated \system{}’s suggestions as significantly more actionable than the original reports.
\end{abstract}

%% file: sections/01_introduction.tex
\section{Introduction}

AI agents can now complete complex tasks across domains, ranging from coding~\cite{chen2025code, ClaudeCode, Codex} to design~\cite{Stitch} to scientific discovery~\cite{lu2024aiscientist, gottweis2025coscientist, jansen2025codescientist}. These advanced agents have shown high-level autonomy in completing complex tasks by breaking them down into smaller steps, and synthesizing all the intermediate results into a coherent output.
However, as their capabilities rapidly grow, user interactions with them remain largely the same traditional request-response paradigm. Most agent systems only \emph{react} to explicit instructions from users before taking action, either in single-turn queries or multi-turn conversations~\cite{ClaudeCode,NotebookLM,Codex}.
The vision of proactive agents that can take initiative to assist users is still not fully realized~\cite{Newendorp2024ApplesKN,horvitz1999principles}; most prior work on proactive agents primarily investigated proactive interactions within short conversation sessions instead of long-horizon interactions~\cite{petridis2026compass}. For example, asking clarification questions~\cite{Patel2025_AR4JSPLP}, getting the user's attention for confirming high-stake steps or fixing agent-failures~\cite{mozannar2025magenticuihumanintheloopagenticsystems,feng2026cocoacoplanningcoexecutionai}, or chiming in a multi-user conversation~\cite{10.1145/3706598.3713749,10.1145/3706598.3713760,wang2025paperping}.

In this paper, we explore an alternative interaction paradigm of supporting proactivity over long-horizon interactions with AI agents. More specifically, we study this in the context of designing a research support agent that observes a researcher's evolving project document, and \emph{proactively} provides literature-grounded suggestions that are \emph{contextualized} to the rich project context. Literature-based research support is well-suited for this exploration due to the increased user interests in leveraging LLMs as research tools \cite{llm_survey,llm_perspectives} and the rich prior work on building literature-based research support capabilities \cite{lee2024paperweaver,kang2022from,Babaian2002_HTDNW9XH}.

The two most common user interaction paradigms for literature-based research support are 1) Querying a deep research system that leverages LLM's reasoning powers to synthesize many retrieved documents; and 2) Recommender systems that proactively send paper recommendations to their users based on a list of seed papers.

Deep research systems (e.g.,~\cite{singh2025scholarQA}) have the potential of leveraging LLMs to understand the user's rich ongoing context. However, the dominant request-response paradigm can be prohibitively laborious for users in two ways.
First, it can be effortful to \emph{compress} the rich context of a complex project down to detailed instructions. Recent findings indicate that, while users increasingly add more ``Complex Contextual Narratives'' to their queries to a literature review AI agent~\cite{Haddad2026UnderstandingUA}, providing context to an AI agent could be more effortful~\cite{petridis2026compass}.
Second, the lengthy and comprehensive output of LLMs can be overwhelming for users, requiring substantial cognitive effort to turn them into concrete \emph{actionable} suggestions---that is, suggestions that propose a specific course of action such as citing a paper for a claim, adopting a baseline, or revising a hypothesis~\cite{graphologue,chopra2023conversational,Kang2023_6GNRRWBF,Sun2024ReviewFlowIS}.

In contrast, paper recommender systems proactively send users recommendations of papers similar to a user's curated seed papers.
While this proactive approach alleviates the burden of constant query formulation, paper recommender systems also have limitations. First,
the static lists of seed papers can quickly go stale as the project's direction and user interests shift~\cite{wu2026rethinkinguserempowermentai}. Prior work has showed that users rarely update their preferences on recommendation systems, leading to stagnated feeds with similar and repetitive recommendations, causing user frustration and abandonment~\cite{wu2026rethinkinguserempowermentai}. Even when a recommended paper is highly useful for a project, \emph{how} it is useful may be buried in a monotonous list of recommendations and the full text of paper, requiring high effort for researchers to realize their usefulness~\cite{kang2022from}. For example, a paper might inform a researcher on how to design their user study, but this can easily be overlooked when receiving a paper recommendation email that typically lists only the titles and abstracts of the recommended papers.

These evidences point to a future where agents observe and leverage a user's actively changing rich context to generate suggestions that are highly relevant and actionable to its users \cite{Newendorp2024ApplesKN,horvitz1999principles}.

In this work we explore \system{}, a system that observes a research project document, tracks the state of a project, formulates interesting and open questions, performs background queries, and expresses the results specifically in the context of the project. Instead of having users provide context at the moment of agent invocation, our approach leverages existing project documents. \system{} regularly checks for changes in the user's project documents, issues timely literature questions based on the project's state to a literature deep research system, and derives actionable suggestions from deep research reports. This approach resolves the problem of stale context often found in recommendation systems, requiring almost no extra effort from the user.

Concretely, this paper makes the following contributions:
\begin{itemize}

    \item Through an iterative design process spanning 15 months, we built three versions of a proactive research assistant, assessed each one with users (Table~\ref{tab:evaluations}), revised the design, and iterated. These experiments yielded the following insights:
     \item Cost of context engineering: Researchers disliked having to invest time and effort crafting the context of their project as background for an agent. Instead, they
     wanted the proactive assistant to monitor their existing artifacts and infer this context (Section~\ref{sec:context-maintenance}).
    \item Cost of contextualization: Instead of a long deep-research style report, users wanted specific, actionable suggestions on how to improve their projects (Section~\ref{sec:actionable-info}).
    \item The final iteration of our system, \system{}, proactively generated agent tasks that participants found timely to their project context (Section~\ref{sec:RQ1-timeliness}), and included tasks they themselves might not have thought of without the system (Section~\ref{sec:rq3-new-ideas}).
    \item These tasks led to suggestions that the participants rated as significantly more \emph{actionable}, \emph{relevant}, and \emph{timely} when users spends similar amount of effort in the baseline condition of reading top sections from a deep research system's output (Section~\ref{sec:RQ2-suggestions}).
\end{itemize}

\begin{table}[t]
  \centering
  \caption{Summary of three evaluations conducted during the iterative design of \system{}.}
  \label{tab:evaluations}
  \footnotesize
  \begin{tabular}{>{\raggedright}p{0.25\columnwidth} >{\raggedright}p{0.50\columnwidth} >{\raggedright\arraybackslash}p{0.11\columnwidth}}
    \toprule
    \textbf{Evaluation} & \textbf{Prototype Tested} & \textbf{Section} \\
    \midrule
    Formative study ($N{=}28$) &
      A technology probe deployment study (8-week): Research interests document with LLM-based paper recommendations via email &
      \S\ref{sec:formative-study} \\
    \addlinespace
    Pipeline Iteration 1 evaluation ($N{=}10$) &
      Initial suggestion pipeline (Iter.~1): evaluating project-state-aware questions and actionable suggestions from project documents &
      \S\ref{sec:pipeline-eval} \\
    \addlinespace
    System evaluation ($N{=}11$) &
      Final \system{}: improved pipeline (Iter.~2) with suggestion-centric and document-centric web interface &
      \S\ref{sec:suggestions-eval} \\
    \bottomrule
  \end{tabular}
\end{table}

%% file: sections/02_related_work.tex
\section{Related work}

Our work draws on two bodies of prior research: (1) proactive AI agents and the types of context they leverage, and (2) literature-based research support tools. We review each below and highlight where existing systems fall short of providing proactive, project-aware assistance.

\subsection{Context for proactive AI Agents}
Research on conversational AI agents has investigated how proactive agents might interact with users. A body of work has focused on identifying the right moment for an agent to chime in during conversation~\cite{Liu2024b_7AS6K8DN,Liu2024a_YMLA762L,Patel2025_AR4JSPLP}. For example, Liu et al. introduce inner thoughts, a mechanism that helps an AI agent determine when to chime in~\cite{Liu2024b_7AS6K8DN}. Other works explore how conversational agents might ask clarifying questions from users to resolve ambiguity and improve performance~\cite{Li2025_RWZI7XAW,Huang2025_6ZQIGHIV,Deng2023a_YDDKJA4W,Tang2025_D4VCW8D7,Zhang2024_FCQKB9UB,Kostric_TUHGEXH8}. The context of these agents and models is mostly conversation history for synchronous interventions.

Other proactive agents embedded in domain-specific tools such as coding~\cite{Pu2025_HLNANI26,Mozannar2024_M75VT844,Chen2024_RAZFVQSN}, searching information~\cite{Heckerman1998_DKR7HTLI} and writing~\cite{Babaian2002_HTDNW9XH,PuertaMelguizo2007_NLVCB8DW,Koskela2018_QXQN98IM} also have limited contexts in terms of scope and time. Some focus more on the appropriate time to offer users assistance without unnecessary interruptions~\cite{Pu2025_HLNANI26,PuertaMelguizo2007_NLVCB8DW}. Systems like Morae determine the decision points and proactively provide relevant task context for users to make informed decisions~\cite{Peng2025MoraePP}. However, these systems focus mostly on synchronous interactions where the users engage actively with the system. The context they used was task-specific and mostly lasted for a short session. \system{}'s context, in contrast, is a long-horizon document for asynchronous proactive interactions.

On the other extreme, there are systems that consume a comprehensive long-horizon multi-modal context~\cite{shaikh2025gum,Shaikh2026LearningNA,Yang2025a_PEE9C5WH}. GUMBO observes a series of unstructured screenshots of a user over time to model users' profiles and provide proactive personalized support~\cite{shaikh2025gum}. Shaikh et al. also introduce LongNAP, a model that predicts user next actions from user multi-modal interaction history~\cite{Shaikh2026LearningNA}. While showing potential in its flexibility, these approaches depend on interaction data which might not give a complete picture of a research project with multiple collaborators contributing from various sources. Further, there is a privacy concern when granting an agent access to all of a user's activities. \system{} leverages existing research artifacts that are more reliable than user interaction history. No prior work has investigated a system that observes a user's documents for a particular project throughout its lifetime.

\subsection{Literature-based Research Support Tools}
There are multiple RAG-based LLM tools designed to support various research tasks from ideation~\cite{pu2025ideasynth,radensky2024scideator}, literature understanding and synthesis~\cite{singh2025scholarQA,feng2026cocoacoplanningcoexecutionai} to paper critique~\cite{afzal2026beyond}.

While useful for their tasks, the scope of context employed by these tools is quite limited. Recent work on paper recommendations has explored how to contextualize recommended papers to a user's context. PaperWeaver uses an LLM pipeline to generate explanations of how a recommended paper is related to a user's paper collection~\cite{lee2024paperweaver}. Kang et al. surface relevance between a recommended paper and a user's publication and interaction history using a knowledge graph~\cite{kang2022from}. The paper collection still needs to be curated manually and the generated explanations are topical instead of actionable suggestions related to a project. Some other tools do not require manual paper curation but have limited context. PaperPing chimes in with related papers during a Slack conversation; however, like other conversational AI agents, its context is limited to prior discussions~\cite{wang2025paperping}. MyScholarQA focuses more on profiling the user based on their publication history~\cite{Balepur2026LanguageMD}, which could be out of sync with their current research projects according to our formative study (Section~\ref{sec:formative-study}). Unlike these systems that focus on a specific research activity, our work considers the context of the whole research project where literature needs shift over time~\cite{soufan2022searching,Inie2022_GEA4QDW6}.

%% file: sections/03_deployment_study.tex
\section{Formative Study with a Technology Probe}
\label{sec:formative-study}

We built an initial technology probe that periodically generated paper recommendations using an LLM based on user's \emph{research interests document}. The recommendations were delivered via email to 28 participants  (4 female and 24 male; NLP, ML, HCI, and CV researchers) over an eight-week period. Using this technology probe, we explored the following questions:

\begin{itemize}
    \item Q1: What types of literature-based information do researchers want from a proactive system?
    \item Q2: Would researchers maintain a free-form document as the primary input to such a system?
    \item Q3: How well could an LLM-based pipeline generate useful recommendations from this richer context?
\end{itemize}

\subsection{Technology Probe: Living Research Interests Document}

The first iteration of our system is a technology probe consisted of three parts: (1)~a shared Google Doc as a \textit{research interests document} that can be freely edited by the participants; (2)~a \textit{paper recommendations pipeline} that periodically read the research interests document, retrieved papers from Semantic Scholar API~\cite{Kinney2023TheSS,singh2023scirepeval} and an open-sourced deep research system~\cite{singh2025scholarQA}, and generated brief explanations of their relevance to the user interests; and (3)~\textit{recommendation emails} delivered twice a week containing the paper recommendations, from which users could add the paper back to their research interests document, asked an AI assistant about them, or gave feedback through the emails. Through the deployment study, we iterated the system based on user's feedback.

\subsubsection{Research interests document}
This served as a lightweight shared memory between the user and the system: users maintained the document as personal notes about their current interests, while the system used it as the source of context for generating recommendations. Users could describe each topic however they wanted (e.g., sentences, bullet lists, links to another documents). The only constraint imposed was that they kept different topics as their own separate sections and added a few relevant papers to each topic. To address the cold-start problem, we seeded each document with the user's five most recent publications under a default ``My papers'' topic.

\subsubsection{Paper recommendations pipeline}
Based on the research interest document, the system generate paper recommendations through an LLM-powered pipeline with the following steps:
\begin{enumerate}
    \item \textbf{Topic and paper extraction} The pipeline fetched the latest content of the user's document and used an LLM to extract the following for each topic: title, description, and associated papers.\footnote{resolved to Semantic Scholar paper IDs} We also allowed participants to include links to other Google Docs, and follow those links to extract additional topics from them. This allowed us to tap into existing documents for their real research projects.
    \item \textbf{Paper retrieval.} The system retrieved paper recommendation using two methods. First, it used the seed papers to find their recent citations and similar papers using the Semantic Scholar APIs~\cite{Kinney2023TheSS}. Second, based on the topic descriptions and notes from the user, the system generated natural language questions to query an open-sourced deep research system and extracted papers from its inline references~\cite{singh2025scholarQA}. Finally, we pooled papers from both methods and ranked them using a recent paper embedding technique~\cite{singh2023scirepeval}, keeping  recommend ones that were most similar to the seed papers.
    \item \textbf{Explanation generation.} For each selected paper, an LLM generated a brief (2--3 sentence) explanation of why the paper was relevant to the user's topic. The prompts took the recommended paper's title and abstract and description of the topic as input (Appendix~\ref{prmpt:formative-exp-topic}). If it cited any of the seed papers, the citing sentence were also included (Appendix~\ref{prmpt:formative-exp-cite}). If the paper was from a deep research output, the question was provided as context (Appendix~\ref{prmpt:formative-exp-question}).
\end{enumerate}

\subsubsection{Recommendation emails}
\begin{figure}[t]
  \centering
  \includegraphics[width=\columnwidth]{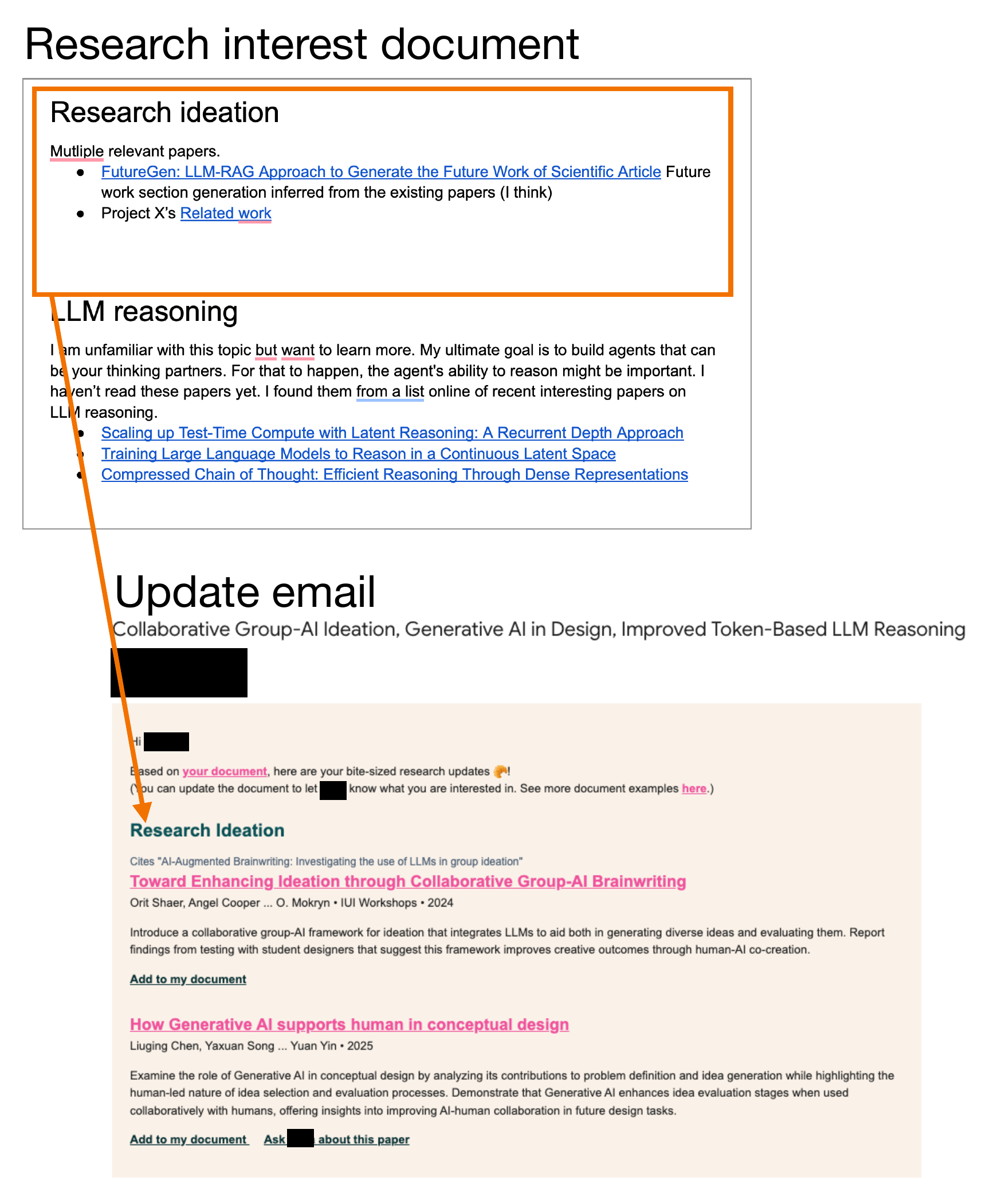}
  \caption{An example of a research update email generated by our technology probe during the formative study. The technology probe extracted topics from a research interest document, retrieved new papers related to the topic, and created explanations of the recommended paper's relationship with the topic.}
  \Description{A screenshot of a research update email showing paper recommendations organized by topics of interest with explanations of relevance.}
  \label{fig:itr1-email}
\end{figure}

The system delivered recommendation emails (Figure~\ref{fig:itr1-email}) twice a week, depending on the availability of highly relevant papers from recent publications.
Each email covered up to two topics, each containing up to two recommended papers. For each recommended paper, the email displayed:
\begin{itemize}
    \item The \textbf{paper title} linked to the corresponding paper details page on Semantic Scholar.
    \item A \textbf{Context label} that includes the topic and how the recommended paper was found based on the above pipeline: \textit{Cites [seed paper title]}) or \textit{Related to [deep research question]}.
    \item A 2--3 sentence \textbf{LLM-generated explanation} of why the paper is relevant to their topic as described in the document.
\end{itemize}
Users could add a paper back to their research interests document by clicking on "Add to my document", allowing it to serve as a seed paper for future recommendations. They could send paper as a context to start  a conversation with an AI reading assistant by clicking ``Ask X about this paper'' button.

\subsection{Procedure}
Participants were informed that they would be receiving research update emails based on their research documents and they could opt out of the study if they wished. We observed their interactions with the research interests document and the emails from February 3rd, 2025 to March 30th, 2025. During the 8-week observation period, the system sent 383 total emails to 28 participants (DP1--DP28).

Over the observation period, we tracked each participant's interaction with the emails and their research interests document. Specifically, we tracked the following events:
\begin{itemize}
    \item \textit{Open paper}: how many times a participant clicked a paper link in the email.
    \item \textit{Add paper to doc}: how many times a participant added a paper from the email to their research interests document.
    \item \textit{Ask agent}: how many times a participant used a paper to start a conversation with the research chat agent.
    \item \textit{\# clicks}: the total number of clicks across all email actions.
    \item \textit{\# doc edits}: the number of edits a participant made to their research interests document.
\end{itemize}

\subsection{Findings}

Overall, the level of engagement was comparable to prior paper recommendation systems in the literature. Each participant received 9--17 emails, and the median click-through rate (CTR) was 9.0\%. For context, the CTR from a prior work on paper recommendation systems was 4.5--5.8\%~\cite{kang2022from}.
Twelve of 28 participants did not trigger any tracked event, though interviews revealed that most had still opened and read the emails. Half of the participants made edits to their interest documents during the 8-week deployment period. With sufficient engagement, we focused our analysis on qualitative insights from the interviews~\cite{braun2006thematic}, with a goal of guiding the design of the next iteration of our system. We reached out to all 28 participants and 13 responded in a timely manner to be interviewed.

\subsubsection{Researchers found the recommendation emails helpful (Q3)}
Eight (of 13) participants indicated that the suggestions were helpful. DP2, DP6, DP17 particularly liked the explanation of paper relevance as they told them what to pay attention to. DP15 spot a few papers that were useful as a citation in their related work.
Some participants with no logged click from the email said that they did read the emails (DP7, DP21). DP7 noted that the recommendations were more ``directly relevant'' than another paper recommendation service they were using. Deep research system was able to find more relevant papers compared to citation- and embedding-based approaches \cite{lee2024paperweaver}. DP21 also said that they found the recommendations from leveraging the deep research system ``starting to be genuinely useful.'' These results shows high opportunities in leveraging existing deep research system for finding highly relevant paper recommendations, combining richer context in an interest document to generate better contextualized explanations.

\subsubsection{Participants wanted actionable insights for their active projects over general interests (Q1)}
\label{sec:actionable-info}

We collected feedback from six participants with low engagement. Participants mentioned that they had read the emails and found the recommendations to be generally interesting, but did not take any actions because they were also not ``super relevant'' to the problem they are solving with their active projects (DP23, DP26).
Contrastly, participants who engaged deeply with the technology probe pointed to how their recommendations were tied to their ongoing projects. Several participants described two tiers of relevance: papers that are ``super relevant'' to a current project---which they would add back to their document---and papers of general interest that they might skim or ignore (DP1, DP26). Some participants (10 out of these 14 that edited their documents) organized their research interests document using active projects on top of topical areas (DP1). DP1 specifically wanted a ``scoop alert'' for new papers that could affect their current project.

Further, beyond discovering relevant papers, participants wanted the system to deliver information they could act upon. The explanation from the probe did not tell them what to do with the information and could have missed important details that they would be interested in at the time. DP7 wanted something ``specific and actionable'' that could enable them to ``close the decision loop''. For example when they would like to see baseline candidates during their experimental design phase. DP23's comment reflected similar sentiment: \textit{``[I want the system to tell] me some information on how to proceed on the project that I am working on.''}
These results suggest participants desire literature-based support to provide \emph{actionable insights} over information that are of high-level interests.

\subsubsection{Maintaining a dedicated research interests document is burdensome (Q2)}
\label{sec:context-maintenance}

We used Google Doc as the primary means for users to communicate their interests to the system.
Participants were split: half of the 28 participants edited their research interests document during the 8-week deployment. Some participants found having an ongoing research interests document helpful for ``organizing ongoing thoughts and threads'' (DP2) or maintaining related work (DP20). Others, like DP23 stated that they were ``too busy and unorganized to put things from [their] head to a maintained document.'' DP17 noted that ``it's a hassle to go back and update the doc.''

Leveraging recent publications to improve cold-start was also insufficient. Multiple participants removed them because it was an ``eclectic collection of things'' (DP17) that no longer reflected their shifting interests (DP3).
These observations suggest that while a dedicated document \emph{can} be valuable, expecting researchers to maintain one specifically for the sake of getting recommendations is unrealistic.

One opportunity emerged from the interviews: the system could draw context from documents researchers already maintain, such as project briefs, weekly meeting slides, or shared notes (DP3, DP7). DP7 suggested that for meeting logs the system should ``just look at the recent one,'' highlighting that the system should handle document staleness gracefully rather than depending on the user's manual curation.

\subsection{Design goals}
Based on insights from the formative study, we formulate the design goals for our proactive literature support system:
 \begin{itemize}
     \item \textbf{DG1}: Help users discover literature relevant to their active research projects instead of general interests.
     \item \textbf{DG2}: Suggest timely and actionable suggestions instead of individual paper recommendations.
     \item \textbf{DG3}: Leverage existing project documents so that users do not need to spend additional effort curating their interests.
 \end{itemize}

%% file: sections/05_system.tex
\section{\system}

\label{sec:system}

Guided by our design goals, we developed \system, a proactive agent that monitors researchers' project documents and delivers actionable, literature-based suggestions contextualized to the project. Instead of asking users to constantly curate up-to-date context, \system uses existing research documents that researchers already maintain ---such as meeting notes or paper drafts (DG3). It then infers the project's specific information needs to surface relevant suggestions (DG1) that are timely and actionable (DG2).

In the next sections, we will first describe the user experience of using \system in an example user scenario. Then we will describe two iterations and evaluation of the system.
\subsection{Example User Scenario}
\label{sec:system-walkthrough}

\subsubsection{Project creation and document upload}
An HCI researcher starts a new research project on ``multi-agent ideation'' with a few research collaborators in her organization. She creates a Google Doc to keep track of the ideas related to this project. The document now contains a description of the project, research questions and a few reference papers. She has already done some literature review in the space but still wants to keep an eye open for relevant work specifically for this project. She adds her Google Doc to a new project on \system, and grants it read permission.

\subsubsection{Initial project details}
Half an hour later, the researcher sees an email in her inbox with a link to \system{}'s dashboard where she can see new suggestions for her project (Fig.~\ref{fig:omakase-system}A). Since this is a new project added to \system{}, she wants to make sure that \system{} correctly inferred information about the project. She opens the \textit{Project details} page~(Fig.~\ref{fig:project-details}).

The system correctly indicates that the project is in an ``Ideation and early experimental design'' stage.
\system{} further explains that it compared recent revisions of the document and found that the project had gone through a few iterations on its framing and pointed out that ``Evaluation criteria'' was described in the document but no results have been added yet.

Even though this inference is correct, she instead wants \system to only focus on supporting ideation so she can explore more broadly. She manually updates the project state to just ``Ideation'', to steer \system{}'s future suggestions.

She also verifies that the system has identified all 17 papers mentioned in her document (Fig.~\ref{fig:project-details}). \system generates short descriptions of how each paper relates to the project. She proceeds to remove two papers that belong to a previous iteration of the project and are no longer relevant.

\subsubsection{Suggestion-centric View}
After inspecting \system{}'s inferred knowledge about the project, the researcher navigates back to the dashboard. She spots one \textit{suggestion card} (Fig.~\ref{fig:omakase-system}A) that points her to ``analogical finding'' literature stating that, based on her multi-agent discussion architecture, she could explore an ``analogical transfer'' mechanism for multi-agent interactions and emphasize that as a contribution to fill the analogical finding/transfer gap. The researcher saves two papers from the suggestions to the project's papers list.

\subsubsection{Document-centric View}
Alternatively, she can also open the \textit{Document view} and see the suggestions in-situ next to highlighted, relevant parts of her document (Fig.~\ref{fig:omakase-system}B). Seeing the document context helps her better understand where each suggestion comes from and identify actions to take related to her original plan. Here, she also sees that a few suggestions point to the missing evaluation section of the document, indicating what needs her attention.

\subsubsection{Adjusting and subscribing to questions}
\begin{figure}[t]
  \centering
  \includegraphics[width=\columnwidth]{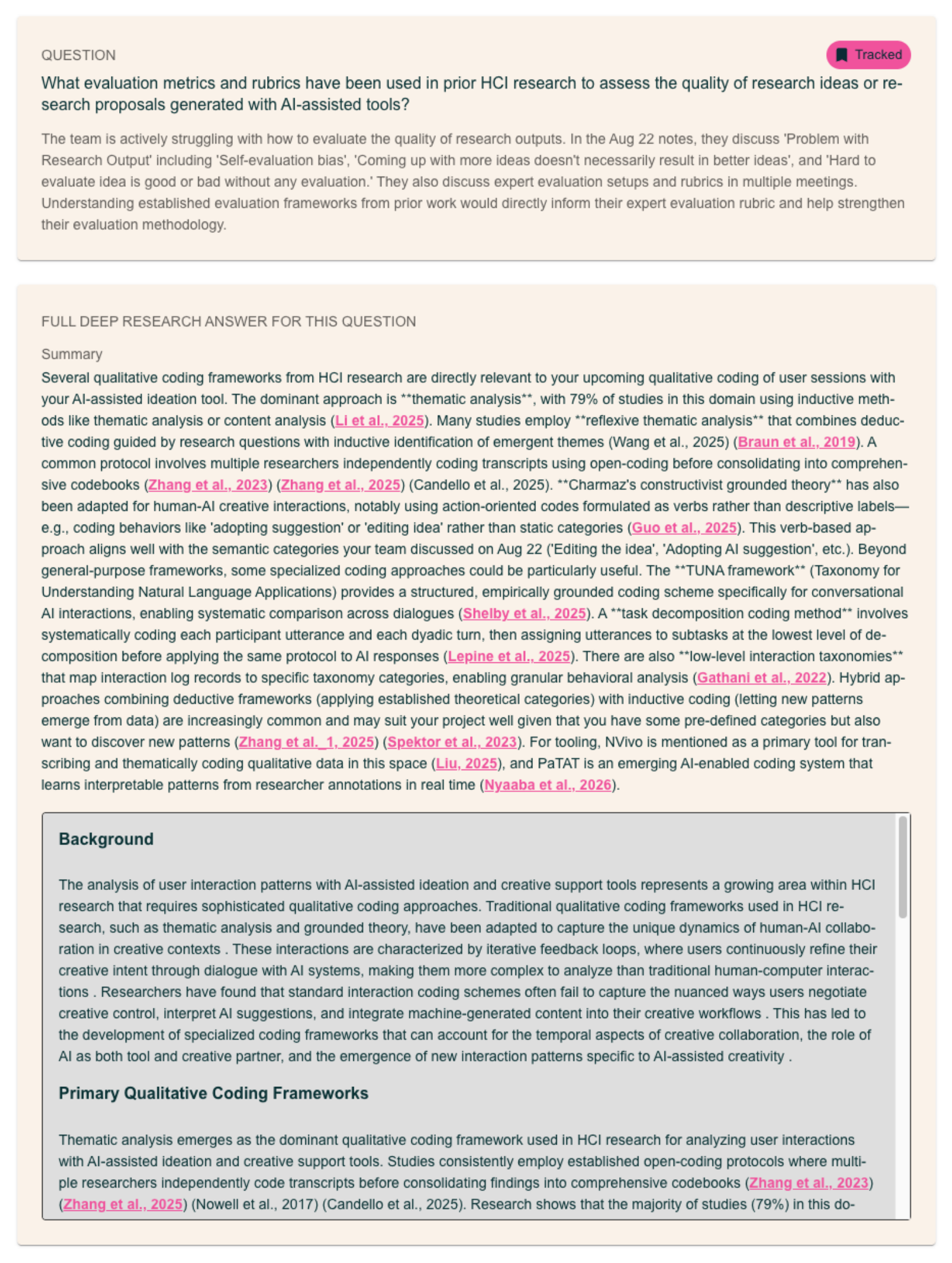}
  \caption{Question page where the user can see the question, rationale for the question, and a summary and the full answers from a deep research system. The user can also view relevant suggestions (Not shown in the figure).}
  \Description{A screenshot of a the question page.}
  \label{fig:question-page}
\end{figure}

\system generates questions to query a deep research system, and contextualizes its output as suggestions. \system exposes this step for transparency and steerability. In the \textit{question page}, she can examine the list of questions as well as add her own question (Fig.~\ref{fig:question-page}).
She can also see the full report of each question before they are contextualized into briefer suggestions in the document-centric and suggestion-centric views.

Further, she saw a question that directly addressed the project's core research question: ``How do existing interactive multi-agent systems allow users to steer, coordinate, and control agent interactions?'' Knowing that she would want to see if there are new answers to this question, she clicked on the ``Track'' button, telling \system to keep track of whether there was any new relevant literature by periodically re-querying the deep research system.

\subsubsection{Ongoing suggestions as the project evolves}
Two months pass. The researcher adjusts \system{}'s update frequency for this project from bi-weekly to weekly to accommodate its weekly meeting schedule (Fig.~\ref{fig:project-details}). As the project moves to the experimental design stage, \system{}'s suggestions shift to focus more on information that could help with the experimental design. The researcher now sees a suggestion to look into metrics to measure the generated ideas from her system. From time to time, she also sees updates to tracked literature review questions when the answer to a question changes.

\subsection{Generation Pipeline Iteration 1}
\label{sec:initial-pipeline}
Once a project document was added to \system, it took the following steps to generate actionable suggestions based on the project context:

\paragraph{Step 1: Document Preprocessing}
\system{} retrieved a document (supported document types are Google Doc, Google Sheet, and Google Slides) from the link provided by the user. It parsed the document into markdown format and extracted a list of potential papers mentioned in the document (Appendix~\ref{prmpt:init-pipeline-paper-extract}). It then retrieved the potential papers metadata (full title) and filtered only those with corresponding papers. \system{} inserted the papers' metadata (Semantic Scholar corpus id and title) to the parsed document.

\paragraph{Step 2: Infer project state and generate questions}
From the document with papers' metadata, \system{} inferred the project state with reasoning. It also extracts relevant questions that could provide information useful for the researcher in the current stage of the project. The prompt provides descriptions of questions a user might be interested in at different stages. It also specifies the model to rank the questions based on relevance and timeliness (Appendix~\ref{prmpt:init-pipeline-doc-analysis}).

\paragraph{Step 3: Issue questions to a deep research system}
\system{} took top questions from the previous step and issued each question to a deep research system~\cite{singh2025scholarQA}. Once the answers were all retrieved the system forwarded the answers to the next step in the pipeline.

\paragraph{Step 4: Generate contextualized suggestions}
Lastly, the suggestions were extracted from deep research responses and within the same prompt ranked by expected usefulness, importance, and relevance to the project phase (Appendix~\ref{prmpt:init-pipeline-suggestion-extraction}). Each suggestion consisted of a text description, a list of mentioned papers, and relevant information from the paper.

\subsection{Generation Pipeline Iteration 1 Evaluation}
\label{sec:pipeline-eval}
To evaluate our initial pipeline, we conducted a study where researchers evaluated the pipeline outputs of their actual project documents. We also gathered researchers' opinions on the cases where project-targeted actionable suggestions were helpful.

\subsubsection{Participants}
We asked 10 research scientists (PP1--PP10) associated with our institution to participate in our study. The participants have different levels of research experience: 2 Ph.D. students, 2 post-doctoral researchers, 1 university professor, and 5 industrial researchers. Their research areas are HCI, NLP, and ML. Each participant provided documents from their ongoing research projects as input to evaluate the pipeline. Seven of the participants also participated in the recommendation emails study.

\subsubsection{Procedure}
We collected 12 research documents from 11 projects. Seven documents were planning and presentation documents that got updated regularly (e.g., for weekly meetings). Two were paper drafts in LaTeX. Three were project proposals or sketches. For each document, we generated the outputs with our pipeline. Each output contained the inferred project state, rationale for the project state inference, and a list of 3-11 questions with 3-8 suggestions per questions. We saved the output of the pipeline in a separate document.

Participants were provided with output document(s) associated with their project(s). If a participant belonged to multiple research projects on our list, they could give feedback on as many documents as they liked. We asked participants to spend 15-20 minutes reading through the suggestions and
provided feedback on the pipeline output and their preferences on receiving updates from the system (see Appendix~\ref{study2:questions} for the full questions list).

\subsubsection{Findings}
\paragraph{The inferred project state is generally correct}
In general, our pipeline inferred the correct state from regularly updated project documents (e.g., meeting notes or weekly presentations) or project proposal in the beginning state.
Six researchers indicated that the inferred project state were correct while four others stated the project state were "close to correct" (PP1) or "Not perfect, but pretty good" (PP4).
Two participants (PP8, PP10) in particular added paper drafts instead of project documents to \system{} also noted that the inferred project states to be generally correct.

\paragraph{Project document-grounded questions and suggestions offers useful information beyond paper recommendations}
In general, participants found useful information in generated suggestions. For example, PP1 found some "really good ideas or suggestions and worth taking a look at". Some also found the suggestions surfaces things they "might not have thought about" even after completing their own literature review (PP2). PP2 and PP8 commented that these questions and suggestions could surface blindspots in the project. In general, participants seeing suggestions based on "[their] meeting notes and researchers notes... could be useful" (PP10).

Some participants noted that the suggestions are more specific and better contextualized than paper recommendations feed, helping them to "know what to pay attention to" (PP6). PP9 commented that the suggestions are not redundant with the existing recommendations. While PP5 "like the document/project-specific nature of this [approach]".

\paragraph{Researchers want control over the generation and delivery of suggestions}
Participants wanted to steer the output by getting more control of the generation process or giving feedback to the system. For example, PP7 wanted to be able to select the questions before issuing them to a deep research system. Participants in general wanted to be able to add their own questions (PP5, PP7, PP8). PP1 wanted to give feedback to the system on the quality of the suggestions. PP3 said that they wanted to edit inferred information and even provided more information to steer the output to fit their needs as their "project doc was not the most comprehensive... it's kind of impossible for the system to know what [they were] working on" (PP3). Unlike real collaborators, the system could not have had background knowledge from past in-person discussion.

Five participants commented that the amount of information presented was too much to view at once. None of them finished reading all suggestions in allotted time. For example, PP7 only got through 2 questions out of 7. However, some had higher information overload threshold than others. For example, PP1 said they would like the information "to be delivered in a way that [wasn't] overwhelming". On the other hand, PP8 said they were fine with the current amount of suggestions as long as they were diverse enough.

Participants also expressed that they wanted control over the timing of suggestion delivery. While some participants would like to receive the suggestions weekly (PP7) or bi-weekly (PP2, PP4), some preferred suggestions being delivered on demand (PP3, PP8, PP9). Some were also interested in the system delivering suggestions as the user makes edits or "when something changes in the environment that affects [their] project" (e.g., new relevant papers got published) (PP8).

The preferred timing varied — and even contradicted. Some wanted suggestions early during brainstorming (PP3), while others explicitly did not want them during ideation when they were already compiling information (PP10). PP10 stated they could "get excited about" receiving suggestions during paper write-up instead. Similarly, some wanted more suggestions at later stages (PP9), while others valued them most during active thinking or paper writing (PP7). This suggests that delivery timing cannot be one-size-fits-all and likely needs to adapt to both project phase and individual preference.

\paragraph{Other improvement opportunities}
Beyond already-mentioned issues, we identified the following opportunities to improve the pipeline and eventual system design:
\begin{itemize}
  \item \textbf{Better output generation.} Some (PP3, PP6, PP7) noted that the questions and suggestions, while generally helpful, sometimes had convoluted phrasing that were hard to parse ("word salad", PP6). PP5 commented that some generated questions should be broken down into sub-questions. This is a common issue with LLM text generation addressable with tighter prompting and post-processing.
  \item \textbf{Awareness of the time of generation.} Some participants wanted to put higher weight on suggestions related to recent actions. They flagged that the output suggestions showed that \system{} did not know that some tasks had already been completed or discarded as directions changed. PP3 noted that the pipeline should be able to infer that their formative had already concluded after 2 months "unless I'm that slow" (PP3). PP7 also noted that the output over-indexed on an abandoned early direction.
  \item \textbf{Suggestions needed to be properly contextualized.} PP3 and PP5 wished the suggestions were more contextualized. Particularly, PP3 wanted the system to explain "where the suggestions were derived". While some other wanted the explanation next to the relevant sections to its corresponding original project documents (e.g., as a comment).
  \item \textbf{Make sure suggestions are non-redundant.} Some suggestions can be redundant with what they already know (PP2), redundant with other suggestions (PP8) or what they are already planning to do (PP3, PP7). While some take this as a signal of knowing the pipeline could be useful, it could be annoying. Further improvement could focus on ensuring all suggestions do not repeat what was already planned in the documents, selecting a set of diverse suggestions and giving users a lightweight way to tell the system what they already know or are planning to do to avoid redundancy.
\end{itemize}

In summary, the evaluation showed that participants were interested in receiving project-document-based suggestions, considered them useful, specific and actionable. The pipeline could infer project state and information need (questions) from active documents that reflected the current state of the project. Participants stated that they want more control over the generation and delivery of suggestions. We took their feedback to improve the pipeline and design of \system.

\subsection{Generation Pipeline Iteration 2}
\label{sec:pipeline-improvements}

Based on the findings above, we made the following improvements to both the generation pipeline and the interface of the final system.

\subsubsection{Pipeline improvements}
Based on the findings from the pipeline evaluation (Section~\ref{sec:pipeline-eval}), we made several improvements to the suggestion generation pipeline addressing the four key areas participants identified: time awareness, output quality, information overload, and redundancy. The final pipeline of \system{} uses Opus 4.6 for all LLM requests in the pipeline (Appendix~\ref{prmpt:omakase-prompts-section}).

\paragraph{Time-aware generation}
PP3 and PP7 noted that the initial pipeline sometimes generated suggestions that overindexed past information. We traced part of this to a document parser bug that dropped date information, causing the model to treat all content as current. After fixing the parser, we also instruct the model to discard open loops that have had no follow-up activity for several weeks (Appendix~\ref{prmpt:omakase-question-extraction}).

\paragraph{Improved output clarity}
Participants reported that some generated questions and suggestions had convoluted phrasing (PP3, PP6, PP7) and that compound questions should be broken into sub-questions (PP5). We updated the underlying model from O3 to Opus 4.6, which produced clearer and more concise text. We also refined the question generation prompt to avoid complicated questions (Appendix~\ref{prmpt:omakase-question-extraction}).

\paragraph{Ranking and selection to manage information overload}
None of the participants in the pipeline evaluation finished reading all generated suggestions within the allotted time, and five explicitly commented that the volume was overwhelming. To address this, we added two explicit ranking stages to the pipeline. First, we added a question selection step that selects a diverse subset of the top questions (Appendix~\ref{prmpt:question-selection}). Second, we added a suggestion selection step that selects the top suggestions based on relevance, actionability, and diversity (Appendix~\ref{prmpt:suggestion-selection}).

\paragraph{Non-redundancy}
Participants noted that some suggestions repeated what they already knew or planned to do (PP2, PP3, PP7), or overlapped with other suggestions (PP8). The suggestion ranking prompt (Appendix~\ref{prmpt:suggestion-selection}) now explicitly instructs the model to penalize suggestions that are redundant with the document's existing plans or with other selected suggestions. Additionally, the system tracks which suggestions a user has previously seen and excludes them from future batches.

\subsubsection{Document viewer}
To better contextualized the suggestions (PP3, PP5), \system{} identifies which part of the document is most relevant to a suggestion (Appendix~\ref{prmpt:document-anchor}). The user can view document section relevant to a suggestion by clicking on the suggestion in the document view (Fig.~\ref{fig:omakase-system}B) to see relevant highlighted text.

\subsubsection{Updates mechanism and control}
\label{sec:heartbeat}
To send participants proactive updates, a background process runs at a configurable interval. For each project, it compares the linked document's last-modified timestamp and the current inferred project state against those recorded at the last pipeline run. If the document has changed or the project state has shifted, the system triggers an \textit{update workflow} that infers the project state and generates new questions to issue them together with tracked questions to a deep research system. The suggestion generation for new questions went as usual. For tracked questions, \system{} compares the new answer with the old answer and only extracts suggestions if there are substantial differences (Appendix~\ref{prmpt:question-diff}).

\system{} gives user control over the timing of update delivery. User can specify how frequent they want \system{} to check for update (from daily to never) (Fig.~\ref{fig:project-details}). They can also manually request new suggestions.

%% file: sections/06_suggestions_evaluation.tex
\section{\system's evaluation}
\label{sec:suggestions-eval}
We conducted a controlled evaluation of \system{}'s Generation Pipeline Iteration 2, focusing on the \textit{quality} of the resulting suggestions, the \textit{timeliness} of the generated questions and the overall user experience using \system{}'s interface. We asked the following:

\begin{itemize}
    \item \textbf{RQ1}: Can \system understand the document to generate timely questions that reflect the current state of the project?
    \item \textbf{RQ2}: How relevant, timely, and actionable do researchers find \system's suggestions compared to directly reading the answers from a deep research system?
    \item \textbf{RQ3}: Does \system help participants discover unexpected information needs that they had not considered?
    \item \textbf{RQ4}: How effortful is it to triage the suggestions?
\end{itemize}

For RQ1, we used \system{} to generate two sets of questions: The first set was based on the up-to-date version of participants' project documents, and the second set was based on an earlier version of the same documents. We then asked participants to compare the two sets of question while being blind to which condition each came from to see if the first set is more timely. For RQ2, we compared \system{}'s suggestions with output of the same deep research report that was used to generate the suggestions, similar to common literature review practices. For RQ3-4, we report qualitative data from post-survey and interview.
\subsection{Participants}
We recruited 11 researchers (SP1--SP11) via social media and academic mailing lists. The participants consisted of 8 Ph.D. students, 1 pre-doctoral researcher, 1 post-doctoral researcher, and 1 research scientist. Two participants had 6-10 years of research experience while the rest has 2-5 years of research experience. None of the participants had participated in the formative study or prior pipeline evaluation.

\subsection{Study protocol}
The study consisted of a pre-study preparation step, followed by an asynchronous evaluation of the system's output and an interview.

\subsubsection{Pre-study: document submissions}
Each participant submitted two copies of their documents from an active research project--a current version and a past snapshot from the document's history.
For each version,  we generated questions that result in 12 deep search reports and 12 \system suggestions.

\subsubsection{Output evaluation}
The output evaluation form had three tasks.

\paragraph{Task 1: Evaluate questions (current vs past)}
Participants received the shuffled list of 23--30 questions, blinded to conditions. For each question, they specified how much they agreed with the statement \textit{``This question is relevant to my project and is a priority to address at its current stage.''} on a 7-point Likert scale (1=``strongly disagree'', 7=``strongly agree''). If a question is not relevant to project at any point in time, they flagged that question to catch the case where the system just generated random questions.

\paragraph{Task 2: Evaluate responses (\system{}'s suggestions vs deep research answer)}
Participants evaluated 12 sets of responses to questions generated from their current document. Because answers from a deep research system can be much longer than a suggestion, we asked participants to select two topics (sections) from each answer that they thought would be most useful. Each set of responses therefore consisted of \system{}'s suggestions and two selected sections from a deep research answer for the same question. We use these two sections to construct two comparison baselines: \textit{Equal effort}, where we compare against the section the participant expected to be most useful, and \textit{Double deep research effort}, where we compare against whichever was rated higher on a given dimension (see below), requiring participants to have read both. Participants evaluated each response in randomized order, blind to condition, rating their agreement on a 7-point Likert scale (1=``strongly disagree'', 7=``strongly agree'') on three dimensions---\textit{relevance}, \textit{actionability}, and \textit{timeliness}:
\begin{itemize}
    \item \textbf{Relevance:} This response is relevant to my project.
    \item \textbf{Actionability:} This response suggests concrete actions I can take (e.g., cite for a claim, adopt a baseline, update a hypothesis).
    \item \textbf{Timeliness:} This response is appropriate for my current stage of work.
\end{itemize}

\paragraph{Task 3: Survey response}
Participants shared their opinions about the questions and the responses in relation to their project. Specifically, we asked if the output helped them think about new ideas or questions (RQ3) and how easy it was to triage the responses (RQ4). See the full list of the survey questions at Appendix~\ref{study3:survey-q}.

\subsubsection{Interview}
We conducted a semi-structured interview with participants. We asked follow-up clarification questions from their evaluation responses. Participants were also presented with \system{} for their current document. We observed how they interacted with the system and asked for their impressions of the system.

\subsection{Analysis}
To test if \system{} generated questions that reflect the priority of the current stage (RQ1), we compared participant's ratings on the relevance and timeliness for questions from the current document and the past document. We aggregated the responses to one score per participant per condition (the median of their ratings) and applied a paired Wilcoxon signed-rank test.

To compare responses from \system{} and a deep research baseline (RQ2), we modeled participants' response ratings using linear mixed-effect models (LMMs).\footnote{We used the \href{https://www.rdocumentation.org/packages/lme4/versions/1.1-38/topics/lmer}{lmer} function from the R lme4 package.} For each of the three rating dimensions, we fitted a crossed random-effects model:
\[
  \text{rating} \sim \text{response\_type} + (1 \mid \text{participant}) + (1 \mid \text{item})
\]
where \textit{response\_type} is a binary factor (deep research vs.\ \system{}); \textit{participant} captures individual participant tendencies, and \textit{item} captures the paired structure of deep research and \system's response originating from the same question for the same participant. We report Holm-corrected $p$-values to account for multiple comparisons.

Deep research systems tend to produce long answers with multiple sections. To keep the length of the deep research and \system{}'s responses comparable, we use only one deep research section for comparison. As introduced above, we consider two baselines. In the \textit{Equal effort} comparison, we use the section the participant expected to be most useful. In the \textit{Double deep research effort} comparison, we use whichever section scored higher on the dimension being tested; this required participants to read both sections, doubling their effort.

\subsection{Results}
\begin{figure}[t]
  \centering
  \includegraphics[width=\columnwidth]{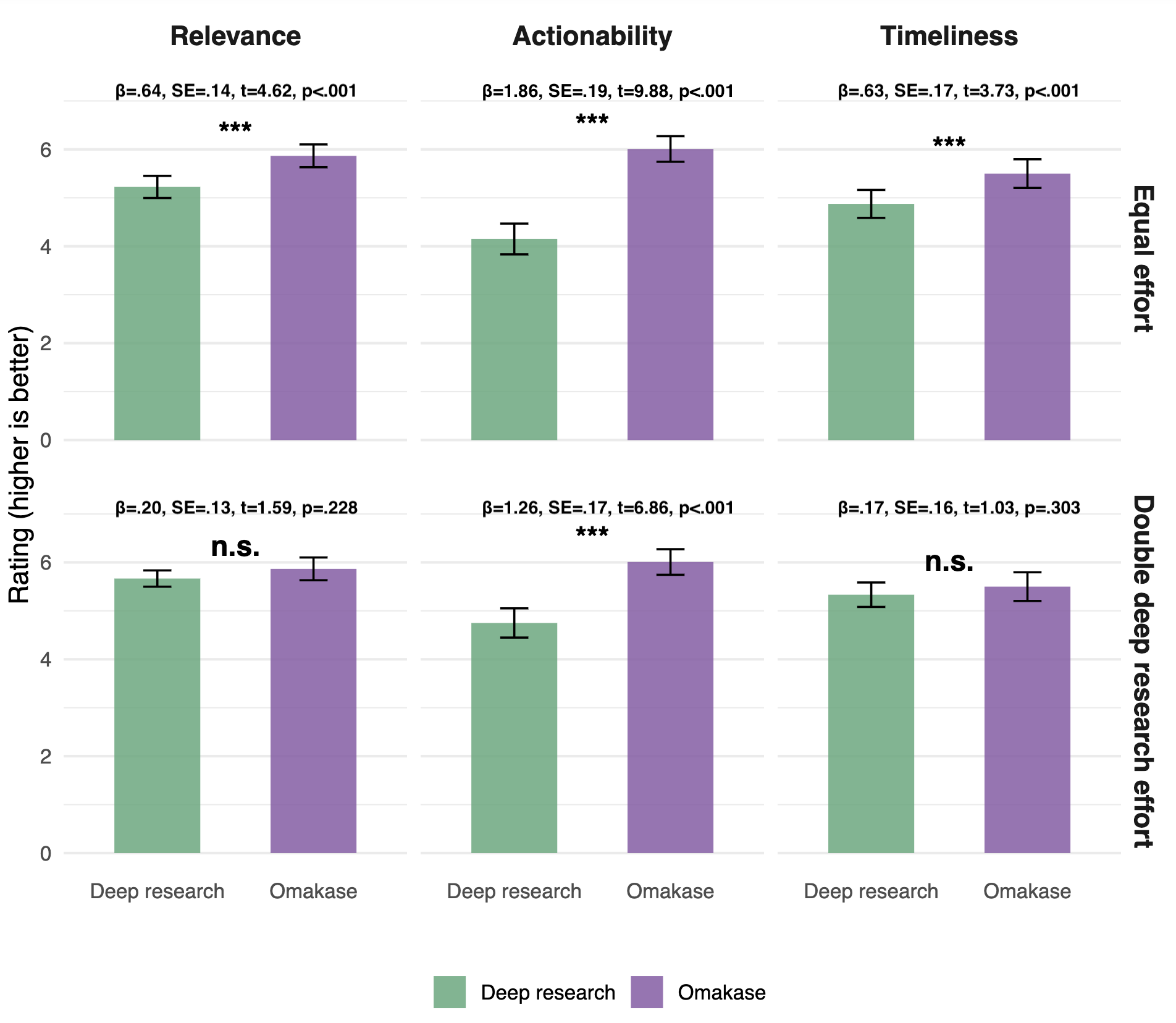}
  \caption{\system{} was rated as more relevant, actionable and timely when compared to a baseline that required similar user effort (i.e., select and read one section from a deep research output). When compared to a stronger baseline where participant spent double the effort (i.e., read two sections), \system{} is still rated significantly more actionable.
  Based on independent LMMs with Holm correction for multiple comparison. {***}${p < .001}$; {n.s.}${p > .05}$}
  \Description{Box plots comparing answer-by-topic and suggestion ratings across relevance, actionability, and timeliness for top-1 and maximum aggregation methods.}
  \label{fig:cross-scale-top1-max}
\end{figure}

\subsubsection{\textbf{Suggestions rated higher than deep research sections (RQ2)}}
\label{sec:RQ2-suggestions}

Based on $N = 240$ observations from 10 participants,\footnote{A participant submitted the documents in the wrong order and was excluded from the responses analysis. Their question ratings were valid after correcting the labels.} we make the following observations:

Comparing \system to when users read similar amount of text in the baseline condition (i.e., selecting one section to read), \system's suggestions were rated significantly higher on relevance ($M=5.87$, $SD=1.32$ vs $M=\text{5.22}$, $SD=\text{1.28}$; $p_{\text{adj}} < .001$), actionability ($M=6.01$, $SD=1.48$ vs $M=4.15$, $SD=1.78$; $p_{\text{adj}} < .001$) and timeliness ($M=5.50$, $SD=1.66$ vs $M=4.88$, $SD=1.61$; $p_{\text{adj}} < .001$; Figure~\ref{fig:cross-scale-top1-max}, top).

We further compared \system to a stronger baseline where users spent double the amount of effort selecting and reading two sections from the deep research output directly.
We found that participants were able to accumulate the information of similar relevance and timeliness. However,
\system{}'s responses were still rated as significantly more actionable than this stronger baseline ($M=6.01$, $SD=1.48$ vs $M=4.75$, $SD=1.69$; $p_{\text{adj}} < .001$; Figure~\ref{fig:cross-scale-top1-max}, bottom).

\subsubsection{\textbf{Questions generated by \system{} from the current stage document aligned more with the current project stage than questions from the past document (RQ1)}}
\label{sec:RQ1-timeliness}
\begin{figure}[t]
  \centering
  \includegraphics[width=0.75\columnwidth]{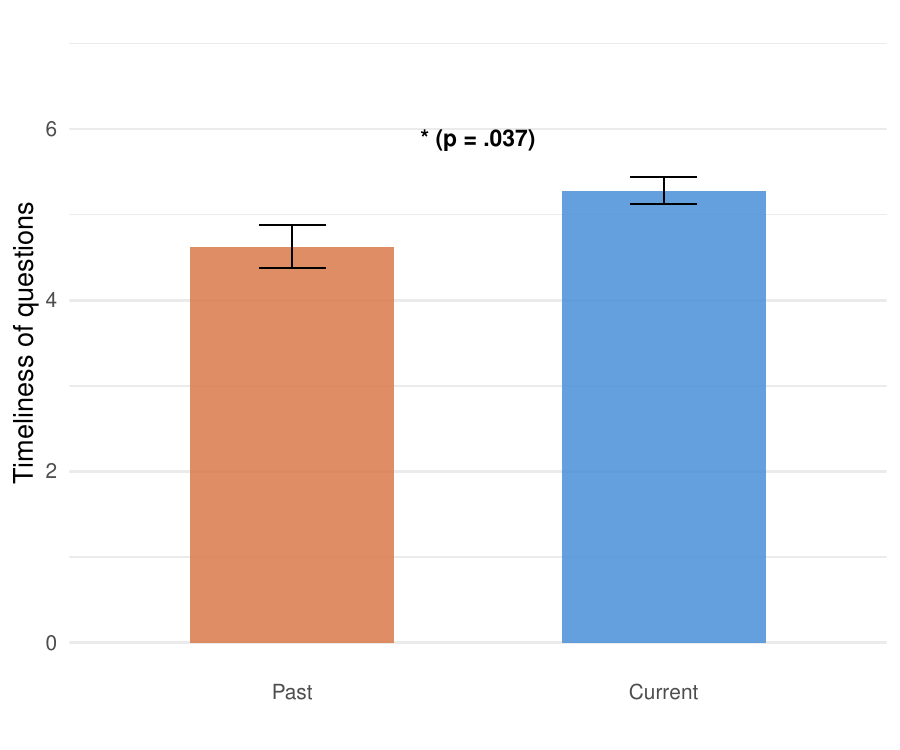}
  \caption{Questions generated from the current document version received significantly higher timeliness ratings than those from the past version (paired Wilcoxon signed-rank test, $V = 57.0$, $p = .037$).}
  \Description{Box plots comparing relevance ratings for questions generated from the current versus past document versions.}
  \label{fig:wilcoxon-question-ratings}
\end{figure}

Across 11 participants,
questions generated from the current version of the document were rated as more aligned with the current stage of project ($M=5.28$, $SE=0.16$) than those from the past version ($M=4.63$, $SE =0.25$; Figure~\ref{fig:wilcoxon-question-ratings}). The difference were statistically significant based on a Wilcoxon signed-rank test ($V = 57.0$, $*p<0.05$).

Flags of questions that were irrelevant to the project at any point in time were rare for both conditions. Out of 11 participants, one participant (SP5) flagged 2 of 15 questions from their past document as irrelevant and one participant (SP6) flagged 2 irrelevant question out of 12 questions.

\system{} generates different sets of questions from projects at different stages, and the questions from the current version of the document better reflect the relevance and priorities of the current stage.

\subsubsection{\textbf{\system{}'s suggestions are easy to triage (RQ4)}}
Nine out of ten participants mentioned it was easy to tell whether the suggestions was worth  reading in details. SP9 commented that \system{}'s suggestions were "short and to the point," making it easy to read in details. In contrast, deep research responses were not as easy to triage. SP7 commented that "A lot more skimming was needed for [deep research responses],... the likelihood of being worth reading is lower (probably 20\%)."
However, two participants (SP6 and SP8) did not find it troublesome to skim through the deep research responses.

\subsubsection{\textbf{\system{}'s questions and suggestions surface new ideas and questions (RQ3)}}
\label{sec:rq3-new-ideas}
Participants reported new ideas and questions inspired by \system{}'s questions and suggestions. SP4, SP5, SP7, and SP10 were made aware of communities and venues that could be closely relevant. Some participants found that the suggestions reminded them to consider experimental design choices that they had not thought of (SP1, SP3, SP10). Some participants also viewed \system{} as feedback giving system (SP1, SP7, SP10). SP1 stated that \system{}'s questions probed for "potential issues" with their project that they hadn't thought of, preventing them from "overclaiming things". SP7 gave an example when \system{} misunderstood ``multi-session''---a key concept in the project---for a ``multi-turn''. That made them realized "how people could misunderstood [their] work". In a way, \system{} can be a mirror that reflects how the work is perceived. SP5 mentioned that having a system "push back on something if there's literature refuting what [they] said" could be helpful.

Some participants also observed that \system{}'s questions were different from the queries they normally issue to their literature tools. SP1 said that the questions were more specific than their ``underspecified queries''. SP5 found the details in \system{}'s questions ``insightful and interesting that [they] wouldn't have framed by [themselves].''

\subsubsection{\textbf{\system{}'s suggestions and deep research answers preference varied}}
Four of ten participants expressed clear preference for \system{}'s suggestions over deep research responses. SP1, SP10 felt that the suggestions were more actionable, similar to how an advisor might provide detailed feedback on their work. While SP2 and SP4 liked how the suggestions referred information from their own research materials. Comparing to a descriptive research answers that were "difficult to map back to what [they] were looking for", the suggestions could "lower the cognitive load of mapping the answers back to the project's context" (SP4).

On the other hand, one participants strongly preferred deep research answers to suggestions. SP6 liked that the deep research answers has more citations details, especially when they were looking for technical depth.

In many cases, participants found that they would prefer one over another for different scenarios. SP5 found that the deep research answers were more helpful when they wanted an overview of a topic. Similarly, SP9 stated that deep research answers was helpful during the initial phases of the project when they are more open to read a lot, while suggestions are appropriate for experimental design and writing phase.

\system{} provides both types of responses to the users. Providing targeted actionable suggestions for those who prefer them while also offering an option to explore more details from the full deep research answers.

\subsubsection{\textbf{Rooms for improvement}}
Despite overall positive reception, participants identified cases where \system{}'s suggestions missed the mark. A recurring theme was that the system sometimes made premature or overconfident inferences about the project's direction. SP7 noted that the system ``sometimes made a `leap' too early and then the suggestions sort of fall flat,'' adding that the deep research answers were ``more careful and sort of said there's no quite anything comparable, which was more accurate.'' SP5 described a case where \system{} overindexed on a topic---notebooks and literate programming---that they had briefly explored but abandoned, treating it as though it were still an active thread of thought.

Relatedly, SP2 was surprised by how often suggestions referenced ideas from the earlier version of their document that no longer appeared in the current draft. They recounted reading suggestions that initially seemed irrelevant, only to realize they traced back to early-stage project plans. SP2 suggested that explicitly noting what is no longer being pursued in the current document could help prevent such misalignment.

%% file: sections/07_discussion.tex
\section{Discussion and Future Work}
\subsection{Granting AI agents access to project document}
There are risks in granting an AI agent access to an ongoing research documents and artifacts~\cite{wagman2025generative}. When we were recruiting participants for the system evaluation, some interested researchers mentioned that their organization prohibited them from sharing their research documents externally. DP16 from the formative study said they did not want to grant an AI agent access to their working research document, citing privacy as the main reason. Indeed, privacy is one barrier to a proactive context-aware agent like Apple's Knowledge Navigator~\cite{Newendorp2024ApplesKN}.

One common feature suggestion for \system{} is to add the capability for an agent to edit or comment on the document directly. However, users might have different preferences in giving comment or write access to their project document to an AI agent and the preferences also depend on the usage context. For example, a user can feel comfortable having an agent editing their document to add a paper to a literature review list but not editing their paper draft. More investigation is needed to understand how to manage AI agents' access to different artifacts in various use cases.

\subsection{Multiple sources of project context}
In this paper, we only explored project context from a single document as an initial exploration step. In practice, a research project can have many artifacts (e.g., project briefs, weekly presentations, paper drafts, experimental results, code, etc.). Multiple sources of project context can give an agent a more complete picture but increase complexity. Future work can explore how AI agents reason about what the user needs from multiple, potentially conflicting, sources.

\subsection{Over-reliance on AI}
\system{}'s suggestions are designed to be easy to triage. Meanwhile, there is concern that using LLM tools could lead to overreliance on AI~\cite{wagman2025generative}. Prior work has shown that AI models are often perceived as unbiased entities~\cite{costello2025just}. The risk is that researchers could accept suggestions from the system without verifying them. One way to mitigate this is by providing evidence links to AI-generated text, making it easier to verify~\cite{krishna2024genaudit}.

\section{Limitations}
While \system{} was designed for long-horizon interactions over the course of a project's lifetime, our evaluation of the system did not cover this aspect. Our user study's attempt to address this issue with two different versions of the document had limited implications. A longer deployment study could provide a more complete picture about the effectiveness of the system.

%% file: sections/08_conclusion.tex
\section{Conclusion}
We presented \system{}, a proactive research assistant that monitors a researcher's evolving project documents and generates literature-grounded, actionable suggestions without requiring explicit queries or manual context curation. Our evaluations showed that \system{} generated suggestions that were significantly more relevant, actionable, and timely than reading equivalent sections of a literature deep research output. Participants also reported that the system surfaced new ideas, relevant communities, and potential weaknesses they had not previously considered, while remaining easy to triage. These findings demonstrate the viability of leveraging existing project artifacts as living context for proactive AI assistance.

%% file: sections/10_appendix.tex
\section{\system{}'s extra details}
\subsection{Implementation details}
We implemented \system{} as a web application. The backend was implemented in Python using Flask for an HTTP server. The front-end was in TypeScript using React framework. We use a Google service account to handle document retrieval and email delivery. \system{} retrieves a paper's metadata from the Semantic Scholar API\footnote{https://api.semanticscholar.org/api-docs} and gets deep research answers from ScholarQA~\cite{singh2025scholarQA}. 

\begin{figure}[H]
  \centering
  \includegraphics[width=\columnwidth]{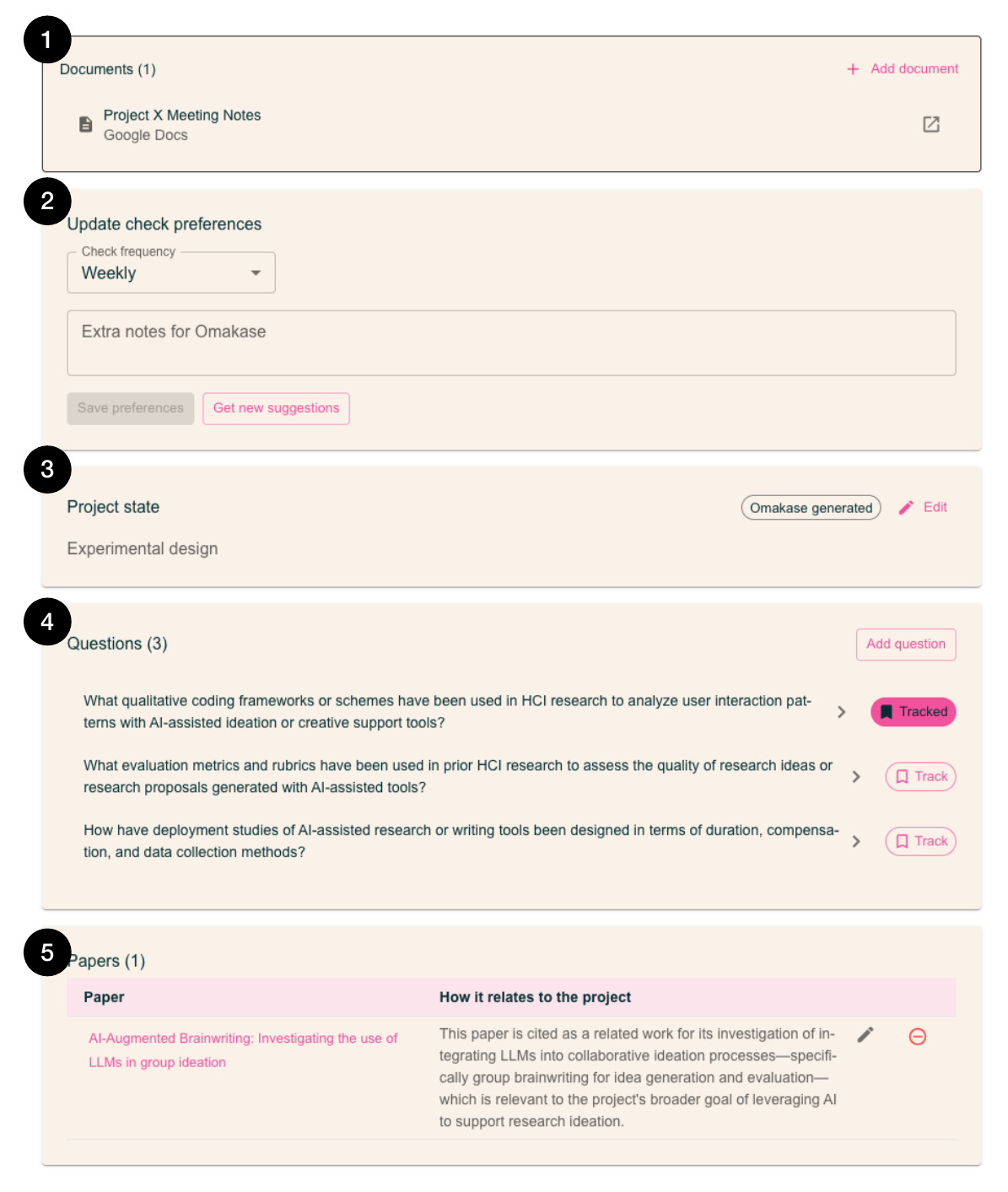}
  \caption{Project details page: The user can edit the inferred project state (3) if \system{} misinterpret. \system{} shows a list of questions for a literature review deep research system that would be helpful for the user(4). The user can add and track questions. There is also a list of papers initially extracted from the document (4); each paper comes with editable explanation of how the paper is relevant to the project. \system{} regularly send user actionable suggestions based on the information in the document. The user can set the frequency of these suggestions or get new suggestions manually (2). }

  \Description{A screenshot of a project detail page.}
  \label{fig:project-details}
\end{figure}
\section{Formative study prompts}
\label{prmpt:formative-prompt}
\subsection{System prompt}
\begin{lstlisting}
You are an intelligent and precise research assistant that can understand
scientific research papers and topics. You are knowledgeable in different
fields and domains of science, in particular, computer science.
You can help a user extract topics from a document in the format of a
Google doc or a markdown file.
\end{lstlisting}

\subsection{Extract papers from a document}
\begin{lstlisting}
Given a document that describes a researcher's interests, return a list of
all mentioned papers in the document.
When possible make sure to provide the full title of the paper.
Include the link to the paper if available.
Always provide the context where the paper is mentioned.
Each paper should either have a link to the paper or the full title of the
paper.
For a paper title, you don't have to be sure that it is a paper title. It
is fine to include the title that are likely to be an academic paper title.
We will look up the paper using the title later.
A paper title is often followed the paper metadata such as the authors, the
abstract, the venue, the url, etc.
Sometimes there is no obvious title, only a link with the url of these
domains: semanticscholar.org, arxiv.org, aclweb.org, acm.org, biorxiv.org.
If so, provide the url and set the text to the label of the link.
Document:
{doc}
\end{lstlisting}

\subsection{Extract topics and questions from a document}
\begin{lstlisting}
Given a document that describes a researcher's research interests and/or
projects, return a list of research topics from the document.
These topics could be research interests, projects, or any other
research-related topics that the researcher is working on or interested in.
The document might have embedded documents that contain additional topics
and papers.
The embedded documents are wrapped between the '===== [document label]
=====' and '===== End [document label] =====' tags.
ALSO EXTRACT the topics and papers from embedded documents and INCLUDE them
in the output. The topic names and descriptions from embedded documents
should take the document label and surrounding context when the document is
mentioned into account.
Each topic should have a short title, a text description that summarizes
the topic or its open questions, a list of relevant papers that are related
to the topic, and a list of questions the researcher can ask about the
topics to find more relevant papers.
The goal of the questions is to help them find papers that help them answer
open questions that they have, fill in the gap of their knowledge, keep
up-to-date with the latest research, or finding supports or contradicting
evidence for their beliefs/claims/hunches.
Make the question specific. Give as much context as possible to help the
user find the right papers.
Make the question focused. Do not try adding more than 2-3 criteria in one
question.
Limit the length of each question to 100 words. For each question, provide
a short rationale that explains how does this question help the user find
relevant and interesting papers.
Explore a few options and pick the most useful 2 questions for each topic.
Make sure that the questions are unique. AVOID questions that are rephrased
versions of existing questions.
The document is the markdown format. Make use of the markdown format
(e.g., headers) to identify the topic.
If a paper doesn't belong to a topic, put it in the closest topic. If a
paper belongs to multiple topics, list it in all relevant topics.
ONLY list the topics and papers mentioned in the document and embedded
documents. DO NOT generate new topics or papers. You may reword some topics
for brevity.
IGNORE any information between and including the lines that contain
---- Template ---- and ----- End Template ----.
Only if there is a group of paper titled 'My Papers' in the doc: group
papers under 'My papers' into a single topic titled, 'My papers'.
Papers are mentioned in <Paper corpus_id="..." /> tags. The corpus_id is a
unique identifier for each paper and is an integer.
Document: {doc}
\end{lstlisting}

\subsection{Generate recommendation explanation with a citation relationship}
\label{prmpt:formative-exp-cite}
\begin{lstlisting}
You are given two research papers: Paper A and Paper B, where Paper A cites
Paper B.
{interest_phases}
Tell me why Paper A would be interesting to me.
Here are titles and abstracts of Paper A and Paper B.

[The Start of Paper A]
Title: {citing_paper_title}
Abstract: {citing_paper_abstract}
[The End of Paper A]

[The Start of Paper B]
Title: {cited_paper_title}
Abstract: {cited_paper_abstract}
[The End of Paper B]

Here is how Paper A cites Paper B:
[Citing sentences in Paper A]
{citing_sentences}
[The End ofciting sentences in Paper A]

Your explanation should be short (only short 2-3 sentences).
DO NOT state what is already obvious from Paper A's title.
Do not mention Paper B if you don't have to. There is no need to repeat
the content of paper B as I have already read it.
KEEP IT BRIEF. Explain what this paper does (propose, introduce, examine,
compare sth etc.) that might be relevant to my interest? Cut out
unnecessary words. I don't want to spend too much time reading this
explanation.
You might rephrase the sentences for readability but DO NOT make up any
information.
Call Paper A by 'this paper'. If needed, call Paper B by their shortened
title and NOT Paper B.
For example, the shortened title of a paper 'GLIDER: Grading LLM
Interactions and Decisions using Explainable Ranking' is 'GLIDER'.
The shortened title of a paper 'Understanding Impact of Human Feedback via
Influence Functions' is 'Understanding Impact...'.
If there is no abstract, no need to mention it. Make sense of whatever you
can from the title.
\end{lstlisting}

\subsection{Generate recommendation explanation with a question}
\label{prmpt:formative-exp-question}
\begin{lstlisting}
{interest_phases}
You are given a research paper.
Tell me information from this paper that helps me answer the question:
{query}.
Here are title and abstract of the paper with relevant text snippet(s)
from the paper.

[The Start of the paper]
Title: {paper_title}
Abstract: {paper_abstract}
Snippet(s): [
{snippets}
]
[The End of the paper]

Your explanation should be short (only short 2-3 sentences).
DO NOT state what is already obvious from the paper's title.
KEEP IT BRIEF. Start with a verb. What does this paper do (propose,
introduce, examine, compare, etc.)? Cut out unnecessary words. I don't
want to spend too much time reading this explanation.
You might rephrase the sentences for readability but DO NOT make up any
information.
If needed, call other referred papers by their shortened title.
For example, the shortened title of a paper 'GLIDER: Grading LLM
Interactions and Decisions using Explainable Ranking' is 'GLIDER'.
The shortened title of a paper 'Understanding Impact of Human Feedback via
Influence Functions' is 'Understanding Impact...'.
\end{lstlisting}

\subsection{Generate recommendation explanation with a topic}
\label{prmpt:formative-exp-topic}
\begin{lstlisting}
{interest_phases}
You are given a research paper.
Tell me the main results and details of this paper that I would find
interesting.
Here are titles and abstracts of the paper.

[The Start of the paper]
Title: {paper_title}
Abstract: {paper_abstract}
[The End of the paper]

Your explanation should be short (only short 2-3 sentences).
DO NOT state what is already obvious from the paper's title.
KEEP IT BRIEF. Explain what this paper does (propose, introduce, examine,
compare sth etc.) that might be relevant to my interest? Cut out
unnecessary words. I don't want to spend too much time reading this
explanation.
I don't want to spend too much time reading this explanation.
You might rephrase the sentences for readability but DO NOT make up any
information.
If needed, call other referred papers by their shortened title.
For example, the shortened title of a paper 'GLIDER: Grading LLM
Interactions and Decisions using Explainable Ranking' is 'GLIDER'.
The shortened title of a paper 'Understanding Impact of Human Feedback via
Influence Functions' is 'Understanding Impact...'.
If there is no abstract, no need to mention it. Make sense of whatever you
can from the title.
\end{lstlisting}

\section{\system{}'s pipeline prompts}
\label{prmpt:omakase-prompts-section}
\subsection{System Prompt}
\label{prompt:system-prompt}
\begin{lstlisting}
You are an intelligent and precise research assistant that can understand scientific research papers and topics. You are knowledgeable in different fields and domains of science, in particular, computer science. You can help a user extract content from a document in the format of a Google doc or a markdown file.
\end{lstlisting}

\subsection{Paper extraction}
\begin{lstlisting}
Given a document that describes a researcher's interests, return a list of all mentioned papers in the document. When possible make sure to provide the full title of the paper.
Include the link to the paper if available.
Always provide the context where the paper is mentioned.
For each paper, if the document explains how the paper relates to the research project (e.g., why it is cited, how it supports or motivates the work, what role it plays), provide a short "project_relation" summary (1-2 sentences). If no such explanation is present, try inferring from the context (surrounding text where the paper is mentioned) or content of the paper. If the paper appears in multiple places, make sure to mention them all. If there is no direct information and inferring the relation is impossible, leave it null.
Each paper should either have a link to the paper or the full title of the paper.
For a paper title, you don't have to be sure that it is a paper title. It is fine to include the title that are likely to be an academic paper title. We will look up the paper using the title later.
A paper title is often followed the paper metadata such as the authors, the abstract, the venue, the url, etc.
Sometimes there is no obvious title, only a link with the url of these domains: semanticscholar.org, arxiv.org, aclweb.org, acm.org, biorxiv.org. If so, provide the url and set the text to the label of the link.
Document:
{doc}
\end{lstlisting}

\subsection{Paper relationships with the project}
\begin{lstlisting}
You are given:
1) A short excerpt from a research project document where a paper is mentioned (or the surrounding context).
2) The paper's title and abstract.
3) An optional "project relation" -- how the author of the document described or implied the paper's relevance to the project.

Write a summary (1-2 sentences) that summarizes how this paper relates to the research project. Use the document context, the stated or implied relation, and the paper's content (title and abstract) to produce a concise, accurate summary. Do not invent relevance; base your summary only on the given inputs. Write in clear, neutral academic language.

Document excerpt (where the paper is mentioned):
{document_excerpt}

Paper title: {paper_title}

Paper abstract:
{paper_abstract}

Project relation (from document, if any): {project_relation}

Summary:

\end{lstlisting}

\subsection{Document analysis and questions generation}
\label{prmpt:omakase-question-extraction}
\begin{lstlisting}
You are an AI research assistant tasked with analyzing a research document of a research project, figuring out the stage the project is in, finding information from existing literature, and providing useful information/insights from existing literature to help the researchers with their research project. Your first task is twofold: first, to infer the current state of the research project, and second, to generate a list of relevant and important questions that can be answered by consulting prior work in the field. The answers to these questions, depending on the stage of the project, should be able to help unblock the researcher, strengthen the quality of the research project, or inform them of unknown related work, etc.

Here is the research document you will be analyzing:

<research_document>
{doc}
</research_document>

First, carefully read and analyze the research document. Then, follow these steps:

1. Infer the project state:
   Based on the content and structure of the document, determine which stage the research project is likely to be in. Possible stages include, but are not limited to:
   - Ideation
   - Literature review
   - Experimental design
   - Data collection
   - Running experiments
   - Data analysis
   - Paper writing

   A research project might have multiple of these activities going on at once.
   Consider factors such as the level of details in different sections, the presence or absence of certain elements (e.g., methodology, results, discussion), and the overall focus of the document.
   If dates are provided, they might be notes of what the researcher was/is focusing on at each point in time. If there are conflicting information, prioritize the recent notes over older ones.
   This step is important as the state of the project determines the type of information the researchers are interested in.

2. Generate relevant questions:
   Create a list of questions that, when answered by consulting prior work, would provide information that is useful for the researcher in the current stage of the project. The question should not be too complicated to accommodate a deep research agent that can't answer complicated questions yet.

   These questions should be:
   a) Relevant: Directly related to the project's topic and goals
   b) Answerable from the literature: Can be addressed by reviewing existing research
   c) Context-independent: Complete in themselves, not requiring additional project context
   d) Timely: Appropriate and useful for the current project stage. Use a good judgement on what types of questions would be good for each research stage from a research scientist's perspective. For example, here are some topics that could be of interested for each state:
   - Ideation: The researcher is coming up with ideas for their project. They would want to know the novelty and feasibility of their ideas, see some inspirations such as solutions from existing papers, use cases, application from different domains. They might not be concerned about details such as experimental protocols or participants' compensation.
   - Literature review: This happens through out the project. Usually in this phase, researchers collect evidences that supports or refutes the claims, relevant claims from prior work, how to situate/frame the project within existing work, think about other work that is related to their ideas, prevent themselves for getting scooped, etc.
   - Experimental design: They are designing a story to learn about something or proving that something they build work. Here they might be looking for baseline ideas, evaluation criteria, experiment setup, things that can inform hypothesis and research question generation/iteration, measures used in prior work, how to plan for statistical analysis
   - Data collection: Usually they are head downs collecting data at this point. Some knowledge about data collection pipeline, method of data collections might be helpful. Also, they might want to know about literature that helps inform their data interpretation.
   - Data analysis: how prior work have analyzed data, what are the conclusions of related work, etc.
   - Paper writing: They are publishing the paper. All written text should be rigorous and thoughtful. Here they might want to know how strong their claims are or about supporting/contrasting evidences to their claims. They would also want to know how strong their framing of the paper is and how to improve them if necessaray.

   Throughout, normal related work are of researcher's interest. Be especially alert on finding work that might scoop this project.

Depending on the stage of the project, good questions could:
   - Help clarify or refine research objectives
   - Identify potential methodologies or experimental designs
   - Suggest relevant baselines or comparison points
   - Address potential limitations or challenges
   - Strengthen the project's theoretical foundation
   - Improve the interpretation or presentation of results
   - etc.,

For each question, provide an explanation of how the knowing the answer to the question could help with the project and point to specific part in the document that helps you come to this conclusion.

3. Prioritize the questions:
   Order the questions from most useful to least useful, considering their potential impact on improving the research project at its current stage.

Present your analysis as structured output with:
- project_state: The inferred project state (a string).
- why_project_state: Briefly explain your reasoning for inferring this project state.
- questions: A list of objects, each with:
  - question: The natural language question (one of the questions ordered from most to least useful).
  - explanation: How knowing the answer could help the project and which part of the document led you to this conclusion.

Order the questions from most useful to least useful. Each question must have both "question" and "explanation" fields.

Ensure that your analysis is thorough, well-reasoned, and directly based on the content of the provided research document.

\end{lstlisting}

\subsection{Question ranking}
\label{prmpt:question-selection}
\begin{lstlisting}
You are helping a research scientist prioritize which questions would be most useful to answer for their current project. The questions will be issued to a Q&A system that retrieves answers from prior work. You will be given information about their project and a list of questions, and you need to rank the questions from most useful to least useful.

Here is the project document:
<project_document>
{doc}
</project_document>

Here is the current state of the project:
<project_state>
{project_state}
</project_state>

Here is the explanation of how this project state was inferred:
<state_explanation>
{why_project_state}
</state_explanation>

Here are the questions to rank:
<questions>
{questions}
</questions>

Your task is to select and rank the top questions that are BOTH highly useful AND diverse. The final set should cover different aspects of the project -- avoid choosing several questions that ask about the same thing or overlap heavily. Favor a mix of questions that address different information needs, methodologies, or angles.

When evaluating usefulness, consider:
- How directly the question relates to the current project goals and methodology
- Whether answering the question would help overcome current limitations or challenges
- How much the answer could advance the project to the next stage
- Whether the question addresses critical gaps in knowledge or methodology
- The potential impact on the project's success if this question were answered

When ensuring diversity, consider:
- Do the selected questions cover distinct topics, sub-questions, or types of information?
- Would answering one question leave another redundant (if so, prefer the stronger one and pick a different question for the set)?
- Does the set collectively give a broad, non-redundant view of what the researcher needs to learn?

Before providing your ranking, use the scratchpad below to think through your analysis:

<scratchpad>
Think through each question and evaluate:
1. How it relates to the current project state and goals
2. What specific value answering it would provide
3. How it compares in importance to the other questions
4. How likely will it be answerable by prior work
5. How it differs from other candidate questions (avoid clustering similar questions in the final set)
</scratchpad>

After your analysis, provide your final ranking. For each question in your ranked list, briefly explain why you placed it at that position and how answering it would benefit the project.

Your final answer should contain only the ranked list of questions (from most useful to least useful) with brief explanations for each ranking decision. The question text should STRICTLY follow the provided question text (if an explanation was provided, do not include the explanation). Do not repeat your scratchpad analysis in the final answer.

\end{lstlisting}

\subsection{Suggestions generation from a deep research report}
\begin{lstlisting}
You are an AI research assistant tasked with generating summary and actionable suggestions based on the following answer for a query issued to a deep research system. Your goal is to give a 1-2 paragraphs summary and provide ways to utilize this information in an ongoing research project described in a given research document.

Here is the question issued to the deep research system:
<question>
{question}
</question>

Why this question was asked for the project (use this to prioritize and anchor your summary and suggestions):
<question_explanation>
{question_explanation}
</question_explanation>

Here is the answer from the deep research system:
<answer>
{answer}
</answer>

The answer includes citation labels that link to academic papers. Here are the labels for this answer:
<citation_labels>
{citation_labels}
</citations_labels>

The research project is described by the following document:
<research_document>
{doc}
</research_document>

The research project is in the "{project_state}" stage because {why_project_state}.

First, your task is to summarize the answer into 2-3 paragraphs. Follow these guidelines:

1. Discard information irrelevant to the question or the project.
2. Prioritize information based on its importance to the project. Important information could:
   - Help clarify or refine research objectives
   - Identify potential methodologies or experimental designs
   - Suggest relevant baselines or comparison points
   - Address potential limitations or challenges
   - Strengthen the project's theoretical foundation
   - Improve the interpretation or presentation of results
3. Avoid repeating information already present in the project document.
4. Highlight new information specifically relevant to the research document (e.g., new papers supporting a claim, findings contradicting assumptions in the document).
5. Use a tone as if one researcher is informing another about relevant points of interest to check out. Avoid prescriptive language.
6. Include the original citation labels of relevant information in the summary.
7. Be specific, providing important details that support the statements.
8. For each statement in the summary, try to include all papers in the original summary that supports that statement.
9. If a paper is mentioned, you MUST use the EXACT original label as provided in citation_labels

Remember, your summary should be concise yet informative, focusing on the most relevant and important information for the research project.

Second, carefully analyze the answer and consider how the information could be applied in various research contexts. Generate a list of actionable suggestions, each accompanied by at least one associated paper and relevant information from that paper.

When formulating your suggestions, consider the following use cases:
1. Strengthening a claim in a research document with more documents
2. Calling for reconsideration of a claim in a research document (in case of conflicting information)
3. Providing ideas for solutions, methods, baselines, datasets, evaluation, or experimental design based on the current stage of a research project
4. Anchor the researcher on "why" this suggestion in the context of their project by pointing back to specific part of the research document.

For each suggestion:
1. Clearly state the suggestion in text and explain how it could improve the project.
2. Provide at least one associated paper label (use all relevant papers if multiple apply).
3. In the suggestions text, mention how the paper(s) inspires the suggestion.
4. Include relevant information from the paper(s) that supports or relates to the suggestion
5. Make sure that the suggestion is timely. (For example, if the project is already in the paper writing stage, it is unreasonable to suggest the researcher to redesign the experiment). Here are some topics that could be of interested for each state:
   - Ideation: The researcher is coming up with ideas for their project. They would want to know the novelty and feasibility of their ideas, see some inspirations such as solutions from existing papers, use cases, application from different domains. They might not be concerned about details such as experimental protocols or participants' compensation.
   - Literature review: This happens through out the project. Usually in this phase, researchers collect evidences that supports or refutes the claims, relevant claims from prior work, how to situate/frame the project within existing work, think about other work that is related to their ideas, prevent themselves for getting scooped, etc.
   - Experimental design: They are designing a story to learn about something or proving that something they build work. Here they might be looking for baseline ideas, evaluation criteria, experiment setup, things that can inform hypothesis and research question generation/iteration, measures used in prior work, how to plan for statistical analysis
   - Data collection: Usually they are head downs collecting data at this point. Some knowledge about data collection pipeline, method of data collections might be helpful. Also, they might want to know about literature that helps inform their data interpretation.
   - Data analysis: how prior work have analyzed data, what are the conclusions of related work, etc.
   - Paper writing: They are publishing the paper. All written text should be rigorous and thoughtful. Here they might want to know how strong their claims are or about supporting/contrasting evidences to their claims. They would also want to know how strong their framing of the paper is and how to improve them if necessaray.

   Throughout, normal related work are of researcher's interest. Be especially alert on finding work that might scoop this project.
6. Use a tone as if one researcher is informing another about relevant points of interest to check out. Avoid prescriptive language.
7. ALWAYS provide citations to any information (e.g., claim, clause, method) that is from a paper with sources.
8. If a paper is mentioned, you MUST use the EXACT original label as provided in citation_labels
9. Provide an anchor to the project document or any understanding of the thing the researcher might be trying to do. Cite specific part of the document that is relevant. The point is to help the researcher see the relevant context quickly. For example, a suggestion could be "If you are still deciding on the evaluation metrics (based on the note from <date>), you could consider using this metrics from paper X which has been used for Y. The pro is ... the con is ..."
10. Create a title that summarizes what the suggestion is in a few words.

If you encounter incomplete information in the answer that requires further investigation, assume you have access to a tool that can retrieve paper abstracts and full texts. Mark such papers with (to_lookup) after the paper label.

Important: Order your suggestions by listing the most useful, important, and relevant to the project phase first. This prioritization will help researchers focus on the most impactful actions.

Before writing your summary and actionable suggestions, use a <scratchpad> to organize your thoughts and identify the most important points to include.

Format your output as follows:
<output>
    <summary>
    [Your 2-3 paragraph summary here, following the guidelines above]
    </summary>

    <suggestions>
    <suggestion>
    <title>[Title that summarizes the suggestion in a few words]</title>
    <text>[Your suggestion here]</text>
    <papers>
    <paper to_lookup=[true or false]>[Paper 1 label] (to lookup - if applicable)</paper>
        ... more papers if applicable ...
    </papers>
    <info>[Relevant information from the paper]</info>
    </suggestion>
    [Repeat for each suggestion]
    </suggestions>
</output>

\end{lstlisting}

\subsection{Suggestion ranking}
\label{prmpt:suggestion-selection}
\begin{lstlisting}
You are helping a research scientist rank recommendations for their project. You need to rank the provided recommendations from most relevant and actionable to least relevant and actionable.

Here is the project documentation:
<project_doc>
{doc}
</project_doc>

Here is the current project state:
<project_state>
{project_state}
</project_state>

Here are the recommendations to rank:
<recommendations>
{recommendations}
</recommendations>

Your task is to rank these recommendations so the top set is BOTH highly relevant/actionable AND diverse. Favor a mix of recommendations that cover different papers, angles, or types of advice -- avoid selecting several suggestions that point to the same paper or give nearly identical advice.

Rank based on three criteria:
1. **Relevance**: How well does the recommendation align with the project goals, current state, and identified information needs?
2. **Actionability**: How much information gain would implementing this recommendation provide? Consider whether the recommendation is specific enough to act upon and likely to yield valuable insights.
3. **Diversity**: Does the selected set cover distinct recommendations? Prefer spreading across different sources, perspectives, and types of suggestions rather than clustering similar ones.

<scratchpad>
Before ranking, analyze each recommendation by:
- Assessing how directly it addresses the project's current needs and goals
- Evaluating the potential information gain and practical value
- Considering the specificity and feasibility of the recommendation
- Noting any dependencies or prerequisites that might affect implementation
- Checking overlap with other recommendations (favor variety: different papers, different angles, different kinds of advice)
</scratchpad>

Provide your analysis and ranking in the following format:

For each recommendation, first provide your reasoning for its ranking position, then list the recommendation. Order them from most relevant and actionable (1st) to least relevant and actionable (last).

**Ranking:**

**1. [Most relevant and actionable]**
Reasoning: [Explain why this recommendation ranks first in terms of relevance to the project and actionability/information gain]
Recommendation: [State the recommendation]

**2. [Second most relevant and actionable]**
Reasoning: [Explain the ranking rationale]
Recommendation: [State the recommendation]

[Continue this pattern for all recommendations]

Make sure to clearly justify each ranking based on relevance, actionability, and (where it helps break ties) how the set stays diverse (different papers, angles, or types of advice).
In the output, the recommendation should be verbatim as the original recommendation text.

\end{lstlisting}

\subsection{Linking document context to a suggestion}
\label{prmpt:document-anchor}
\begin{lstlisting}
You are an AI scientific research assistant that links research suggestions to specific sentences in a research document.

Today's date is {today}.

Below is a research document split into numbered sentences:

<document_sentences>
{document_sentences}
</document_sentences>

Below is a list of suggestions that were generated for this research project. Each has a unique ID, text content, and optionally an explanation of why this suggestion was made:

<items>
{items}
</items>

Your task: For each suggestion, identify the single sentence in the document that it is most relevant to. Use the suggestion's explanation (when provided) to understand the intent behind the suggestion and find the best document sentence.

Guidelines:
1. Only link to sentences that are clearly relevant -- do not force matches.
2. Prefer specific, meaningful sentences over generic ones (e.g., section headers alone are usually not enough).
3. Copy the exact verbatim sentence text as the quote.
4. For each item, return exactly one match -- the single sentence that most directly and importantly relates to the suggestion. When evaluating whether a sentence is relevant, consider a few surrounding sentences as context (the sentences immediately before and after it), not just the candidate sentence in isolation. If multiple nearby sentences are relevant, choose the one with the strongest connection to the suggestion. Prefer recent content over older content. Each document has its own ordering; try to infer where the recent content is (could be at the end, the beginning, or the middle). Make use of dates when given, and use today's date to determine what is recent. If no sentence is clearly relevant, return an empty matches list.
5. The sentence_index must correspond to the number shown before each sentence in the document.
6. For each match, provide a "reasoning" field: 1-2 sentences explaining why this document sentence is relevant to the suggestion. Be specific about the connection.
7. For each match, provide a "location" field: a short label describing where in the document this sentence appears. Use the nearest section heading, tab name, or date if available (e.g., "Experiment Design > 22 Aug 2025", "Introduction", "Notes from Week 3"). If no clear heading exists, describe the position (e.g., "Near the beginning", "Midway through the document").

Return your answer as a JSON object with this structure:
{
  "<item_id>": {
    "matches": [
      { "sentence_index": <int>, "quote": "<exact sentence text>", "reasoning": "<why this sentence is relevant>", "location": "<where in the document>" }
    ]
  }
}

Include every item_id from the input, even if matches is empty.

\end{lstlisting}

\subsection{Comparing old deep research answers with new answers}
\label{prmpt:question-diff}
\begin{lstlisting}
    You are an expert research assistant comparing two ScholarQA answers to the same question, generated at different times. Your goal is to identify substantive differences that could lead to actionable suggestions for the researcher.

## Question
{question}

## Project State
{project_state}

## Previous Answer
{old_answer}

## Latest Answer
{new_answer}

## Instructions

Compare the two answers and determine whether there are meaningful differences relevant to the project. Meaningful differences include:
- New papers or findings not present in the previous answer
- Contradicted or updated findings
- New methods, tools, or approaches that have appeared
- Shifts in consensus or emphasis within the field

Do NOT flag differences that are:
- Merely rephrased or reorganized versions of the same content
- Minor wording changes without substantive new information
- Differences in citation formatting or paper ordering

For each meaningful difference, produce a concise, actionable suggestion the researcher can use.

Return a JSON object with:
- "has_meaningful_diff": true/false
- "suggestions": a list of objects, each with "title" (short headline), "text" (actionable suggestion), and "info" (supporting context from the new answer)

If there are no meaningful differences, return has_meaningful_diff=false and an empty suggestions list.

\end{lstlisting}

\section{Initial pipeline prompts}
\label{init-pipeline-prompt}
\subsection{System prompt}
Same as \ref{prompt:system-prompt}.

\subsection{Paper extraction}
\label{prmpt:init-pipeline-paper-extract}
\begin{lstlisting}
Given a document that describes a researcher's interests, return a list of all mentioned papers in the document. When possible make sure to provide the full title of the paper.
Include the link to the paper if available.
Always provide the context where the paper is mentioned.
Each paper should either have a link to the paper or the full title of the paper.
For a paper title, you don't have to be sure that it is a paper title. It is fine to include the title that are likely to be an academic paper title. We will look up the paper using the title later.
A paper title is often followed the paper metadata such as the authors, the abstract, the venue, the url, etc.
Sometimes there is no obvious title, only a link with the url of these domains: semanticscholar.org, arxiv.org, aclweb.org, acm.org, biorxiv.org. If so, provide the url and set the text to the label of the link.
Document:
{doc}

\end{lstlisting}

\subsection{Document analysis and question generation}
\label{prmpt:init-pipeline-doc-analysis}
\begin{lstlisting}
You are an AI research assistant tasked with analyzing a research document of a research project, figuring out the stage is in, finding information from existing literature, and providing useful information/insights from existing literature to help the researchers with their research project. Your first task is twofold: first, to infer the current state of the research project, and second, to generate a list of relevant and important questions that can be answered by consulting prior work in the field. The answers to these questions should be able to help unblock the researcher, strengthen the quality of the research project, or inform them of unknown related work, etc.,  depending on the stage of the project.

Here is the research document you will be analyzing:

<research_document>
{doc}
</research_document>

First, carefully read and analyze the research document. Then, follow these steps:

1. Infer the project state:
   Based on the content and structure of the document, determine which stage the research project is likely to be in. Possible stages include, but are not limited to:
   - Ideation
   - Literature review
   - Experimental design
   - Data collection
   - Running experiments
   - Data analysis
   - Paper writing

   A research project might have multiple of these activities going on at once.
   Consider factors such as the level of details in different sections, the presence or absence of certain elements (e.g., methodology, results, discussion), and the overall focus of the document.
   If dates are provided, they might be notes of what the researcher was/is focusing on at each point in time. If there are conflicting information, prioritize the recent notes over older ones.
   This step is important as the state of the project determines the type of information the researchers are interested in.

2. Generate relevant questions:
   Create a list of questions that, when answered by consulting prior work, would provide information that is useful for the researcher in the current stage of the project. These questions should be:
   a) Relevant: Directly related to the project's topic and goals
   b) Answerable from the literature: Can be addressed by reviewing existing research
   c) Context-independent: Complete in themselves, not requiring additional project context
   d) Timely: Appropriate and useful for the current project stage. Use a good judgement on what types of questions would be good for each research stage from a research scientist's perspective. For example, here are some topics that could be of interested for each state:
   - Ideation: The researcher is coming up with ideas for their project. They would want to know the novelty and feasibility of their ideas, see some inspirations such as solutions from existing papers, use cases, application from different domains. They might not be concerned about details such as experimental protocols or participants' compensation.
   - Literature review: This happens through out the project. Usually in this phase, researchers collect evidences that supports or refutes the claims, relevant claims from prior work, how to situate/frame the project within existing work, think about other work that is related to their ideas, prevent themselves for getting scooped, etc.
   - Experimental design: They are designing a story to learn about something or proving that something they build work. Here they might be looking for baseline ideas, evaluation criteria, experiment setup, things that can inform hypothesis and research question generation/iteration, measures used in prior work, how to plan for statistical analysis
   - Data collection: Usually they are head downs collecting data at this point. Some knowledge about data collection pipeline, method of data collections might be helpful. Also, they might want to know about literature that helps inform their data interpretation.
   - Data analysis: how prior work have analyzed data, what are the conclusions of related work, etc.
   - Paper writing: They are publishing the paper. All written text should be rigorous and thoughtful. Here they might want to know how strong their claims are or about supporting/contrasting evidences to their claims. They would also want to know how strong their framing of the paper is and how to improve them if necessaray.

   Throughout, normal related work are of researcher's interest. Be especially alert on finding work that might scoop this project.

Depending on the stage of the project, good questions could:
   - Help clarify or refine research objectives
   - Identify potential methodologies or experimental designs
   - Suggest relevant baselines or comparison points
   - Address potential limitations or challenges
   - Strengthen the project's theoretical foundation
   - Improve the interpretation or presentation of results
   - etc.,

3. Prioritize the questions:
   Order the questions from most useful to least useful, considering their potential impact on improving the research project at its current stage.

Present your analysis in the following format:

<analysis>
<project_state>
[Inferred project state]
</project_state>

<reasoning>
[Briefly explain your reasoning for inferring this project state]
</reasoning>

<questions>
1. [Most useful question]
2. [Second most useful question]
3. [Third most useful question]
...
[Continue listing questions in order of decreasing usefulness]
</questions>
</analysis>

Ensure that your analysis is thorough, well-reasoned, and directly based on the content of the provided research document.

\end{lstlisting}

\subsection{Suggestion generation}
\label{prmpt:init-pipeline-suggestion-extraction}
\begin{lstlisting}
You are an AI research assistant tasked with generating summary and actionable suggestions based on the following answer for a query issued to a deep research system. Your goal is to give a 1-2 paragarphs summary and provide ways to utilize this information in ongoing research project described in a research document below.

Here is the question issued to the deep research system:
<question>
{question}
</question>

Here is the answer from the deep research system:
<answer>
{answer}
</answer>

The answer includes citation labels that link to academic papers. Here are the labels for this answer:
<citation_labels>
{citation_labels}
</citations_labels>

The research project is described by the following document:
<research_document>
{doc}
</research_document>

The research project is in the "{project_state}" stage because {why_project_state}.

First, your task is to summarize the answer into 2-3 paragraphs. Follow these guidelines:

1. Discard information irrelevant to the question or the project.
2. Prioritize information based on its importance to the project. Important information could:
   - Help clarify or refine research objectives
   - Identify potential methodologies or experimental designs
   - Suggest relevant baselines or comparison points
   - Address potential limitations or challenges
   - Strengthen the project's theoretical foundation
   - Improve the interpretation or presentation of results
3. Avoid repeating information already present in the project document.
4. Highlight new information specifically relevant to the research document (e.g., new papers supporting a claim, findings contradicting assumptions in the document).
5. Use a tone as if one researcher is informing another about relevant points of interest to check out. Avoid prescriptive language.
6. Include the original citation labels of relevant information in the summary.
7. Be specific, providing important details that support the statements.
8. For each statement in the summary, try to include all papers in the original summary that supports that statement.
9. If a paper is mentioned, you MUST use the original label as provided the citation_labels

Remember, your summary should be concise yet informative, focusing on the most relevant and important information for the research project.

Second, carefully analyze the answer and consider how the information could be applied in various research contexts. Generate a list of actionable suggestions, each accompanied by at least one associated paper and relevant information from that paper.

When formulating your suggestions, consider the following use cases:
1. Strengthening a claim in a research document with more document
2. Calling for reconsideration of a claim in a research document (in case of conflicting information)
3. Providing ideas for solutions, methods, baselines, datasets, evaluation, or experimental design based on the current stage of a research project

For each suggestion:
1. Clearly state the suggestion in text and explain how it could improve the project.
2. Provide at least one associated paper label (use all relevant papers if multiple apply).
3. In the suggestions text, mention how the paper(s) inspires the suggestion.
4. Include relevant information from the paper(s) that supports or relates to the suggestion
5. Make sure that the suggestion is timely. (For example, if the project is already in the paper writing stage, it is unreasonable to suggest the researcher to redesign the experiment)
6. Use a tone as if one researcher is informing another about relevant points of interest to check out. Avoid prescriptive language.
7. If a paper is mentioned, you MUST use the original label as provided in citation_labels

If you encounter incomplete information in the answer that requires further investigation, assume you have access to a tool that can retrieve paper abstracts and full texts. Mark such papers with (to_lookup) after the paper label.

Important: Order your suggestions by listing the most useful, important, and relevant to the project phase first. This prioritization will help researchers focus on the most impactful actions.

Before writing your summary and actionable suggestions, use a <scratchpad> to organize your thoughts and identify the most important points to include.

Format your output as follows:
<output>
    <summary>
    [Your 2-3 paragraph summary here, following the guidelines above]
    </summary>

    <suggestions>
    <suggestion>
    <text>[Your suggestion here]</text>
    <papers>
    <paper to_lookup=[true or false]>[Paper 1 label] (to lookup - if applicable)</paper>
        ... more papers if applicable ...
    </papers>
    <info>[Relevant information from the paper]</info>
    </suggestion>
    [Repeat for each suggestion]
    </suggestions>
</output>

\end{lstlisting}

\section{Output from existing conversational AI assistant with RAG (January 2026)}
We investigated existing agentic systems to see how well they could extract project-specific suggestions from a project document. We created two versions of a project document from a finished project historical versions. One from a few weeks after the project started and the other from a few weeks before the paper was submitted. We created a prompt for each version and sent the prompt to ChatGPT\footnote{https://chatgpt.com/} and Gemini\footnote{https://gemini.google.com/app} in January 2026. These systems were chosen as strong RAG-enabled conversational AI assistants. See the prompts used below.

We found that the suggestions responses were quite generic (see example outputs below). One suggestions from ChatGPT state that the researchers should conduct a pilot experiment for an HCI research project at the experimental design phase. While the paper surfaced works well for the initial of the project, these systems failed to surface more targeted specific papers that were typically helpful at later stages of the project. Further, there was one instance where the Gemini's responses contained hallucinated a paper. 

\label{prompt:chat-baseline}
\subsection{Paper recommendations prompt}
\begin{lstlisting}
Here is a document from my research project. Help me find the
most relevant papers to this project that have not been mentioned in the document.
Each paper should have a title, a url to the paper, a short explanation of why
it is relevant to my project, and the place in the document where the paper will be relevant.

Here is the document:
----- Start of Document -----
{doc_content}

----- End of Document -----
Provide your response in a clear, structured format
\end{lstlisting}

\subsection{Literature-grounded suggestion prompt}
\begin{lstlisting}
Here is a document from my research project. Give me literature-grounded suggestions
on things to do reach a successful conclusion to my project.
Each suggestion should have a text description, related paper(s) that backed the suggestions, a url for each paper, a short explanation of why
it is a good suggestion, and the place in the document where the suggestions will be relevant.

Here is the document:
----- Start of Document -----
{doc_content}

----- End of Document -----
Provide your response in a clear, structured format.
\end{lstlisting}

\subsection{Paper recommendations response example (Gemini)}
\begin{lstlisting}
   Gemini said
   Based on your research notes from May to August 2024, your project focuses on a mixed-initiative, node-based system that helps researchers transition from fragmented ideas to grounded research proposals using LLM agents and literature synthesis.

   While your document mentions several key systems like CoQuest, PaperWeaver, and Luminate, there are critical academic works in the "Human-AI Co-Creativity" and "AI for Science" (AI4Sci) domains that would provide a theoretical and technical backbone for your specific features.

   Recommended Research Papers
   Paper Title	URL	Key Relevance to Your Project	Where it fits in your Document
   Corgi: Content-based Roadmap Generation for Scientific Literature	https://arxiv.org/abs/2210.15041	Explores how to generate structured "roadmaps" from papers. It directly informs how your "Node Canvas" can represent a literature space.	Aug 12 & July 30: Sections regarding the tree structure of nodes and "Node-based Lit review."
   ChatEval: Towards Better LLM-based Evaluators through Multi-agent Debate	https://arxiv.org/abs/2308.07201	This paper details how multiple LLM personas can debate to reach a higher-quality evaluation than a single agent.	June 4 & May 30: Discussions on using "Opponent critiques" and "Agent by persona" for convergence.
   ResearchAgent: Iterative Hypothesis Generation via Knowledge Graph Reasoning	https://arxiv.org/abs/2404.07738	Focuses on using LLMs to iteratively refine research hypotheses by connecting disparate papers.	Aug 6 & July 11: Notes on how to generate "3-5 research ideas" and "grounding" them in literature.
   Managing Design Conflict in Human-AI Co-creation	https://dl.acm.org/doi/10.1145/3544548.3581105	Analyzes how users negotiate agency when an AI suggests changes---crucial for your "User Control/Agency" metrics.	Aug 15 and July 29: User survey questions regarding agency, trust, and AI suggestion quality.
   Detailed Analysis & Integration
   1. Structured Sensemaking (The "Node Canvas")
   Your project heavily emphasizes a canvas-based workflow where nodes represent research components. Corgi is a highly relevant predecessor because it moves beyond list-based paper recommendations (like Perplexity) and focuses on the relationship between findings.

   Integration Tip: In your July 30 notes, you mention "Semantic Zoom" and "Heuristics for what nodes to include." The Corgi paper provides a framework for how to visually cluster papers based on "problem-solving" trajectories, which could help automate your node-linking logic.

   2. Multi-Agent Critique & Debate
   You discussed having an "opponent" persona to critique ideas on June 4. ChatEval provides a formal methodology for this. It suggests that having agents with distinct roles (e.g., a "Methodologist" vs. a "Domain Expert") improves the reliability of the evaluation.

   Relevance to Aug 15: Use the debate-style evaluation to help generate the "Actionable Feedback" you want to provide to users in their proposals.

   3. Human-AI Agency and Trust
   In your final notes on Aug 15, you mention measuring NASA-TLX and User Control/Agency. The paper "Managing Design Conflict" is vital here because it highlights that users often feel "stolen from" if the AI generates too much of the idea.

   Relevance to Aug 7: This paper explains why your users might want the AI to "not change existing content too much" and helps frame your qualitative interview questions about "perceived quality" vs. "perceived ownership."

   Suggested Next Step
   Would you like me to draft a set of interview questions specifically based on the "Human-AI Co-creation" literature to help you measure the User Control and Agency metrics you identified in your August 15th notes?
\end{lstlisting}
\subsection{Suggestions response example (ChatGPT)}
\begin{lstlisting}
   1) De-risk the user study by running structured usability inspections + a pilot before you recruit ``real'' participants

   What to do (actionable):

    Run a 60-90 minute usability inspection pass  (heuristic evaluation + cognitive walkthrough) on the end-to-end ``happy path'' workflow (seed papers --> first node --> lit review panel use --> proposal generation --> export).
    Convert findings into a short, prioritized ``must-fix before study'' list focused on: broken flows, unclear action labels, UI overload, and latency/error states.
    Run at least 1-2 pilots with full logging + recording, and treat the pilot as a test of both software stability and study protocol timing.

   Related papers (with URLs):

   Jakob Nielsen (1994). Usability Inspection Methods --- https://static.aminer.org/pdf/PDF/000/089/680/usability_inspection_methods.pdf
   Wharton et al. (1994). The Cognitive Walkthrough Method: A Practitioner's Guide --- https://www.colorado.edu/ics/sites/default/files/attached-files/93-07.pdf
   van Teijlingen & Hundley (2001). The importance of pilot studies --- https://sru.soc.surrey.ac.uk/SRU35.PDF

   Why this is a good suggestion:
   Inspection methods reliably surface breakdowns that will otherwise dominate user behavior (and your outcomes), while pilot studies increase the likelihood your main study succeeds by catching protocol + instrumentation failures early.

   Where this is relevant in your document:

   Aug 12 / Aug 15: long bug list; ``Fix bugs and system''; ``Record the pilot study''
   Jul 29: pilot feedback indicates confusion and unclear signals for AI suggestions
   2) Lock a single ``intended workflow'' and implement progressive disclosure to reduce overwhelm

   What to do (actionable):

   Write a one-page ``intended workflow spec'' (the steps users should follow; what each panel is for; when AI intervenes).
   Redesign UI to progressively reveal functionality: initial modal --> minimal node actions --> only then advanced node types/edge operations.
   Add ``what to do next'' micro-prompts tied to each major action (e.g., after adding papers, recommend: ``Pick one node to refine'' rather than offering the whole menu).

   Related papers (with URLs):

   Ben Shneiderman (2002). Creativity support tools --- https://creativetech.mat.ucsb.edu/readings/schneiderman_creativitysupport.pdf

   Why this is a good suggestion:
   Creativity-support and exploratory tools routinely fail when users face too many undifferentiated options; a staged workflow preserves exploration while lowering cognitive overhead and increasing adoption of the ``core loop.''

   Where this is relevant in your document:

   Aug 6: ``Envisioned workflow'' (seed papers + idea --> nodes --> proposal panel)
   Aug 12 / Jul 23: ``Not sure what the intended workflow should be''; ``Floating New node types are confusing''
   Aug 6 / Aug 7: ``Progressively show functionality''; ``Semantic zoom''
   3) Make ``grounding in literature'' inspectable: every key claim should link to evidence and be easy to verify

   What to do (actionable):

   For lit review panel + node suggestions, attach per-claim citations (paper + section/snippet) and show them on demand (hover/expand).
   Add a ``show why'' view: what evidence was used, what was ignored, and confidence/coverage indicators (e.g., ``2/5 papers mention this finding'').
   Add a fast ``open paper'' affordance everywhere you reference a paper (fix missing links).

   Related papers (with URLs):

   Tintarev & Masthoff (2007). Evaluating the effectiveness of explanations for recommender systems --- https://scispace.com/pdf/evaluating-the-effectiveness-of-explanations-for-recommender-9rlhua1nou.pdf

   Why this is a good suggestion:
   Explanations are most useful when users can evaluate and act on them; making evidence visible supports trust calibration and reduces the risk that users treat fluent summaries as authoritative.

   Where this is relevant in your document:

   Aug 12: ``No links to the papers in paper collection''
   Jul 23: ``Highlight certain phrases with link to paper''; ``How to tie statements to literature''
   Aug 5 / Aug 6: ``Make paper references in node suggestions connected to LitReviewPanel''
   4) Align success metrics with your claims using validated instruments (not bespoke questions only)

   What to do (actionable):

   Decide your top-level claims (e.g., ``improves idea quality,'' ``improves exploration diversity,'' ``reduces workload,'' ``increases trust/agency'').
   Map each claim to: (a) a validated scale, (b) an expert rating dimension, and (c) one behavioral/log proxy.
   Keep bespoke questions, but use them as diagnostic, not primary outcome measures.

   Related papers (with URLs):

   Jian, Bisantz, & Drury (2000). Foundations for an Empirically Determined Scale of Trust in Automated Systems --- https://www.researchgate.net/profile/Ann-Bisantz/publication/247502831_Foundations_for_an_Empirically_Determined_Scale_of_Trust_in_Automated_Systems/links/54465e450cf2f14fb80f3405/Foundations-for-an-Empirically-Determined-Scale-of-Trust-in-Automated-Systems.pdf
   Cherry & Latulipe (2014). Quantifying the Creativity Support of Digital Tools through the Creativity Support Index --- https://mkremins.github.io/refs/CSI.pdf

   Why this is a good suggestion:
   Validated scales improve interpretability and defensibility (reviewers can recognize constructs like trust/creativity support), and they reduce ``metric hacking'' risk late in the project.

   Where this is relevant in your document:

   Aug 15: ``User Control/Agency, user trust''; ``User survey''; ``Rate AI suggestions quality... Specificity... Actionable''
   Jun 11 / Aug 6: ``subjective satisfaction''; ``expert rating''; ``coherence/diversity''
   5) Strengthen expert rubric validity by adding calibration + agreement measurement (and designing for disagreement)

   What to do (actionable):

   Create 6-10 ``anchor'' proposals and have experts rate them together once; document the rationale for scores.
   During main scoring, compute inter-rater agreement and track drift; if agreement is low, use the anchors to re-calibrate.
   Predefine how you will aggregate scores (mean vs. median vs. discussion-based resolution) and what to do when experts disagree.

   Related papers (with URLs):

   Krippendorff (2011). Computing Krippendorff's Alpha-Reliability --- https://www.asc.upenn.edu/sites/default/files/2021-03/Computing%20Krippendorff%27s%20Alpha-Reliability.pdf
   Cohen (1960). A Coefficient of Agreement for Nominal Scales --- https://journals.sagepub.com/doi/10.1177/001316446002000104
   Amabile (1982). The Social Psychology of Creativity: A Consensual Assessment Technique --- https://www.hbs.edu/faculty/Pages/item.aspx?num=7355

   Why this is a good suggestion:
   Your document already anticipates a rubric-driven expert evaluation; adding agreement measurement and calibration turns ``expert ratings'' into a rigorous measurement instrument rather than anecdotal judgments.

   Where this is relevant in your document:

   Aug 15: ``Expert rubric... see agreement, then adjust score''; ``Fix expert review rubric... procedures''
   Aug 13: ``Established key expert evaluation metrics, come up with examples''
   6) Standardize study artifacts to remove confounds (format/length/content visibility)

   What to do (actionable):

   Enforce format parity: both conditions produce proposals with the same section structure, max length, citation formatting, and export format.
   Add ``generate button''/LLM help in baseline only if you can justify it as a controlled comparison; otherwise keep baseline minimal but fair (time-on-task and feature access must be defensible).
   Build a ``proposal indistinguishability checklist'' for the experimenter to verify before each session.

   Related papers (with URLs):

   van Teijlingen & Hundley (2001). The importance of pilot studies (use pilots to validate parity + protocol timing) --- https://sru.soc.surrey.ac.uk/SRU35.PDF

   Why this is a good suggestion:
   Most null/ambiguous results in comparative HCI studies come from uncontrolled differences in outputs and participant effort; format parity makes your findings interpretable and publishable.

   Where this is relevant in your document:

   Aug 13: ``maintain same format and length... suggest length ~3 sentences''
   Aug 12: ``proposals indistinguishable in format from both conditions''
   7) Turn your logging plan into a mechanism story (not just ``we logged clicks'')

   What to do (actionable):

   Define a compact event schema tied to hypotheses (e.g., ``divergence events,'' ``convergence edits,'' ``literature verification opens,'' ``AI regeneration requests,'' ``node linking operations'').
   Predefine 5-8 process metrics (e.g., exploration breadth = #distinct ideas/nodes; convergence depth = #revision cycles on selected nodes; verification behavior = evidence opens per claim).
   Use interviews to validate the meaning of log patterns (``I opened papers because...'').

   Related papers (with URLs):

   Ericsson & Simon (1980). Verbal Reports as Data --- https://acs.ist.psu.edu/ist597/papers/ericssonS80.pdf
   Braun & Clarke (2006). Using thematic analysis in psychology --- https://educationaldevelopment.uams.edu/wp-content/uploads/sites/57/2025/01/9-Thematic_analysis.pdf

   Why this is a good suggestion:
   Protocol-style verbalization and thematic analysis let you interpret behavioral traces correctly (logs alone can be misleading), producing a credible explanation of how your system changes ideation---not only whether it does.

   Where this is relevant in your document:

   Aug 15: ``Write the script to pull the logs''
   Aug 7 / Jul 7: ``Log user actions... litreview panel and other places''; ``Data Logging system for user study''
   8) Make interview questions explicitly about strategy adoption and ``teachable workflows,'' then analyze them rigorously

   What to do (actionable):

   Add protocol items that ask participants to narrate: (a) what they tried first, (b) what they ignored and why, (c) what made them trust/distrust outputs, (d) what they would teach a colleague.
   Code interviews with a stable codebook (strategy types; breakdown points; trust calibration behaviors; convergence triggers).
   If multiple coders, compute reliability for key codes.

   Related papers (with URLs):

   Braun & Clarke (2006). Using thematic analysis in psychology --- https://educationaldevelopment.uams.edu/wp-content/uploads/sites/57/2025/01/9-Thematic_analysis.pdf
   Krippendorff (2011). Computing Krippendorff's Alpha-Reliability --- https://www.asc.upenn.edu/sites/default/files/2021-03/Computing%20Krippendorff%27s%20Alpha-Reliability.pdf

   Why this is a good suggestion:
   Your notes already emphasize ``strategies of using the system'' and ``how to teach someone''; rigorous qualitative analysis turns those insights into publishable design implications and grounded claims.

   Where this is relevant in your document:

   Aug 15: ``What are the strategies... if you were to teach someone''; ``Interview questions... noticeable instances''
   Jul 19: ``What's the next step after seeing the AI suggestion''
   9) Define ``better research ideas'' with explicit dimensions and measure them consistently across conditions

   What to do (actionable):

   Decide which dimensions you want to improve (e.g., novelty, variety/diversity, feasibility, coherence, specificity, actionability).
   Use a combination of: (a) objective-ish ideation metrics (where applicable), (b) expert ratings (with CAT-style approach for creativity/novelty if desired), and (c) participant self-assessment.
   Ensure the system UI nudges toward these dimensions (e.g., dedicated ``feasibility check'' node type; ``comparison'' view for alternatives; ``next-step experiment sketch'' prompt).

   Related papers (with URLs):

   Shah, Vargas-Hernandez, & Smith (2003). Metrics for measuring ideation effectiveness --- https://web.mit.edu/sloan2/dese/readings/week02/shahHernandezSmith03.pdf
   Amabile (1982). The Social Psychology of Creativity: A Consensual Assessment Technique --- https://www.hbs.edu/faculty/Pages/item.aspx?num=7355
   Shneiderman (2002). Creativity support tools --- https://creativetech.mat.ucsb.edu/readings/schneiderman_creativitysupport.pdf
   Cherry & Latulipe (2014). Creativity Support Index (CSI) --- https://mkremins.github.io/refs/CSI.pdf

   Why this is a good suggestion:
   Without an explicit, consistent definition of ``quality,'' you risk measuring proxies that don't reflect the project's value (or that vary by topic). These instruments give you a coherent measurement story tied to creativity/ideation theory and CST evaluation practice.

   Where this is relevant in your document:

   Aug 6: ``Measures: completeness, exploration diversity, coherence... novelty and impact''
   Jun 18 / Jun 13: divergence vs. convergence; ``fleshing out idea is prereq for evaluating''

   If you want, I can convert this into a single ``execution checklist'' that maps (a) each study claim --> (b) system features needed --> (c) instruments/logs/interview items --> (d) analysis outputs, using only constructs already present in your notes.
\end{lstlisting}

\section{Questions for pipeline evaluation}
\label{study2:questions}
\begin{itemize}
    \item Is the inferred project state correct?
    \item What do you like/dislike about the current output format (answers to questions and suggestions)?
    \item Where would you like the information like this to be delivered (e.g., on the doc, email, slac, etc)?
    \item How does this type of information compare to other literature updates you receive (e.g., S2 feed)? Is it redundant?
    \item What kind of feedback would you give to Omakase to improve its output?
    \item Do you want to receive an update like this for this document in the future? (If so, how regularly do you want them?)
\end{itemize}
Beyond these questions, participants were encouraged to provide free-form comments. They could also comment directly on the output document.

\section{Output evaluation survey questions}
\label{study3:survey-q}
\begin{enumerate}
    \item Which three responses were the most useful for you? For each, specify the question number and whether it was a suggestion or an answer by topic, and explain why.
    \item Have any questions, suggestions and answers by topic help you think about new ideas or questions that you haven't thought of before? If so, please explain.
    \item How did the suggestions and answers by topic differ in usefulness for your project? In what ways, if any, were they similar?
    \item How easy was it to tell whether a suggestion was worth reading in detail?
    \item How easy was it to tell whether an answer by topic was worth reading in detail?
    \item Were any suggestions interesting but not directly related to your project's current stage? If so, which ones and why were they interesting?
    \item Were any suggestions very similar to what you already planned to do? If so, which ones? Could someone else reading your document have come to the conclusion that you already planned to do these things?
    \item Did you notice any suggestions that overlapped or repeated each other? If so, which ones?
    \item Did any suggestions make incorrect assumptions about your project? If so, which ones? Was the correct information available in your document, or what additional context would have been needed?
  
\end{enumerate}